\documentclass[11pt,a4aper]{article}%
\pdfoutput=1
\usepackage{jheppub}
\usepackage{rotating}
\usepackage{array}
\usepackage{amsmath}
\usepackage{slashed,hyperref}
\usepackage{adjustbox}
\usepackage{color}
\usepackage{epstopdf}
\usepackage{overpic}
\usepackage{enumerate}

\usepackage[normalem]{ulem}
\definecolor{orange}{RGB}{255,127,0}

\newcommand{\ewXL}{$ew\chi L$}

\DeclareMathOperator{\Tr}{Tr}

\begin{document} 

\preprint{
\begin{flushright}
IPPP/18/11\\
IFIC/17-30\\
FERMILAB-PUB-18-241-T\\
\end{flushright}}

\title{Signals of the electroweak phase transition at colliders and gravitational wave observatories}

\author[a]{Mikael Chala,}
\author[b,c]{Claudius Krause,}
\author[d,e]{and Germano Nardini}

\affiliation[a]{Institute of Particle Physics Phenomenology, Physics Department, Durham University, Durham DH1 3LE, UK}
\affiliation[b]{IFIC, Universitat de Val\`encia-CSIC, Apt. Correus 22085, E-46071 Val\`encia, Spain}
\affiliation[c]{Theoretical Physics Department, Fermi National Accelerator Laboratory, Batavia, IL, 60510, USA}
\affiliation[d]{Albert Einstein Center, Institute for Theoretical Physics, University of Bern, Sidlerstrasse 5, CH-3012 Bern, Switzerland}
\affiliation[e]{Faculty of Science and Technology, University of Stavanger, N-4036 Stavanger, Norway}

\emailAdd{mikael.chala@durham.ac.uk}
\emailAdd{ckrause@fnal.gov}
\emailAdd{nardini@itp.unibe.ch}


\abstract{
If the electroweak phase transition (EWPT) is of strongly first order due to higher dimensional operators, the scale of new physics generating 
them is at the TeV scale or below.
In this case the effective-field theory (EFT) neglecting operators of dimension higher than six may overlook terms that are  relevant for the EWPT analysis.
In this article we study the EWPT in the EFT to dimension eight. We estimate the reach of the future gravitational wave observatory LISA for probing the region  in which the EWPT is strongly first order and compare it with the capabilities of the Higgs measurements via double-Higgs production at current and future colliders. 
We also match 
different UV models to the previously mentioned dimension-eight EFT
and demonstrate that, from the top-down point of view, the double-Higgs production is not the best signal to explore these scenarios.}

\maketitle

\section{Introduction}
We accurately measured the Higgs mass and its couplings to the heavy SM fermions and gauge bosons~\cite{Khachatryan:2016vau}, but the ElecroWeak (EW) sector remains very uncertain. Within the current constraints, there is still room for a vast variety of phenomena that exhibit intriguing signatures. One of them is the possibility that the Higgs field produces gravitational waves  when it acquires a Vacuum Expectation Value (VEV)~\cite{Witten:1984rs, Hogan:1986qda, Kamionkowski:1993fg}.
For this to happen, the EW Symmetry Breaking (EWSB) must proceed via a Strong First Order EW Phase Transition (SFOEWPT). This is only possible if physics beyond the Standard Model (SM) exists, as such a transition requires the finite-temperature Higgs potential to behave radically differently from the one of the SM~\cite{Kajantie:1996mn, Rummukainen:1998as, Laine:1998jb, Csikor:1998eu}~\footnote{This different behaviour is not needed in (peculiar) setups where the EWPT is preceded by some exotic phenomena. One example is the warped extradimension framework in which the EWPT is forbidden till when the decomposite-composite transition starts~\cite{Creminelli:2001th, Randall:2006py, Nardini:2007me, Konstandin:2010cd}. A further case occurs when inflation has a reheating temperature below the EW scale~\cite{GarciaBellido:1999sv, Krauss:1999ng, GarciaBellido:2002aj, Smit:2002yg, Konstandin:2011ds, Arunasalam:2017ajm, vonHarling:2017yew}.}.

Numerous extensions of the SM exhibiting a SFOEWPT have been considered in the literature. In most of the cases, the main ingredient to depart from the SM finite-temperature Higgs potential is to invoke new light particles in the thermal plasma coupled to the Higgs~\cite{Quiros:1999jp, Morrissey:2012db}. In general, making these new light fields naturally compatible with the present LHC constraints requires to rely on either extra symmetries or particular parameter regions. The strategies to test  these scenarios are therefore very model dependent.
However, new light particles are not a necessary ingredient to achieve a SFOEWPT. Higher-dimensional operators, obtained by integrating out heavy fields, can also provide large non-SM contributions to the Higgs potential. In this case, the lack of observation of additional particles would not be ascribed to circumstantial conditions, but simply to a considerable gap between the EW scale and the new physics scale, $f$. 

At the EW scale, the theory with $\mathcal O(f)$-mass fields  can be described by an effective Lagrangian containing the SM interactions as well as a tower of effective operators suppressed by powers of $1/f$.  Among these operators, the interactions $\mathcal O_n=(\phi^\dagger \phi)^{\frac{n}{2}}$ have a radical impact on the Higgs potential (here $\phi$ is the Higgs EW doublet and $n$ an even integer larger than four). 
Refs.~\cite{Zhang:1992fs, Grojean:2004xa, Bodeker:2004ws, Delaunay:2007wb, Grinstein:2008qi, Damgaard:2015con, deVries:2017ncy} studied in detail the dynamics of the EWPT in the presence of only $\mathcal O_6$. They showed that, in order for the EWSB to proceed via a SFOEWPT, the new physics scale must be $f\lesssim 600$ 
GeV if its couplings are of order one. The small gap between the EW scale $v\sim 246$ GeV and the required $f$ carries two major implications. \textit{(i)} It points out that the EFT to dimension six is inaccurate. Any observable related to the EWPT receives corrections of order $\sim v^2/f^2\gtrsim 20\%$. The next tower of effective interactions, namely $\mathcal O_8$, must be included. \textit{(ii)} It triggers the question of which new physics, at a scale of few hundreds 
GeV can produce such large modifications of the Higgs potential without being constrained by other Higgs measurements or direct LHC searches. We address these two points in this paper.

Thus, in section~\ref{sec:ewpt}, we present the analysis of the EWPT in this extended EFT. We investigate
the validity of the mean-field approximation. Moreover, we accurately determine the regions of the parameter space leading to the SFOEWPT, and characterize the consequent  gravitational wave spectrum.
We also identify the precise values of the coefficients of  $\mathcal O_6$ and $\mathcal O_8$ that the future gravitational wave observatory LISA can test. Finally, we compare this region with the one that can be tested at colliders, sensitive to  $\mathcal O_6$ and $\mathcal O_8$ via the Higgs self coupling measurements.

Next, in section~\ref{sec:weakly}, we discuss those models that can be matched to the EFT above without conflicting with current data. Among the most natural candidates, we single out a weakly-coupled custodial quadruplet extension.
We study its phenomenology and find that at the LHC the most promising search for such an extension is to look for multi-lepton signals.
Section \ref{sec:conclusions} is devoted to our conclusions.

\section{The electroweak phase transition in the EFT to dimension eight}\label{sec:ewpt}
Let us consider the SM extended with the effective operators $\mathcal O_6$ and $\mathcal O_8$, the relevant Lagrangian being
\begin{equation}
 L = L_\text{SM} + \frac{c_6}{f^2}(\phi^\dagger\phi)^3 + \frac{c_8}{f^4}(\phi^\dagger\phi)^4~,
\end{equation}
where $L_\text{SM}$ is the SM Lagrangian, $\phi$ is the Higgs doublet and $f$ stands for the scale of new physics.
In this section we determine the VEV of the Higgs at the critical and nucleation temperatures, $v_{T_c}$ and $v_{T_n}$,  the latent heat of the phase transition, $\alpha$, and the inverse duration time of the phase transition, $\beta/H$, in this non-minimal EFT.
The results we obtain extend those previously obtained in the literature (see \textit{e.g.} Refs.~\cite{Grojean:2004xa, Delaunay:2007wb}), where only $\mathcal O_{6}$ has been considered (despite the low cutoff and the consequent potential breaking of the EFT approach). 
\subsection{Finite temperature potential}\label{sec:finite}
The first ingredient we need is the Coleman-Weinberg effective potential at finite temperature; see Ref.~\cite{Quiros:1999jp} for a review. 
In the Landau gauge and in the $\overline{MS}$ renormalization scheme, the one-loop effective potential $V_{1\ell}$ of our EFT scenario can be expressed as
\begin{eqnarray}
\label{eq:V1ell}
V_{1\ell} = V_{\rm tree}+\Delta V_{1\ell} ~,
\end{eqnarray}
with
\begin{eqnarray}
\label{eq:Vtree}
V_{\rm tree} &=& -\frac{\mu^2}{2} h_c^2 +\frac{\lambda}{4} h_c^4 +\frac{c_6}{8 f^2} h_c^6+\frac{c_{8}}{16f^{4}} h_c^8~,\\
\Delta V_{1\ell}&=&\Delta V_{1\ell, T=0} + V_{1\ell,T\ne 0}~,\\ \label{eq:VT0}
\Delta V_{1\ell, T=0} &=& \sum_{i=h,\chi,W,Z,t}\frac{n_i\, m_i^2(h_c)}{64 \pi^2} \left( \log\frac{m_i^4(h_c)}{v^2} - C_i \right)~,\\
V_{1\ell,T\ne 0} &= & \frac{n_t\, T^4}{2 \pi^2} J_f\left(m_t^2(h_c)/T^2\right)
+
\sum_{i=h,\chi,W,Z}
 \frac{n_i\, T^4}{2 \pi^2} J_b\left(m_i^2(h_c)/T^2\right) 
~,\label{eq:VT}
\label{eq:CWpot}
\end{eqnarray}
where $\Delta V_{1\ell, T=0}$ is the temperature-independent one-loop contribution and  $V_{1\ell,T\ne 0}$ is the (one-loop) remaining part. The variable $h_c$ is a constant background field of the Higgs. In Eq.~\eqref{eq:VT0}, $n_i$ are the degrees of freedom $n_W=2 n_Z= 2 n_\chi = 6 n_h = -n_t/2 =6$, while $C_i$ is equal to 5/6 for gauge bosons and 3/2 for scalars and fermions. The $h_c$-dependent squared masses $m_i^2$ are
\begin{eqnarray}
m_h^2(h_c) &=& -\mu^2 + 3 \lambda\, h_c^2 + \frac{15 c_6}{4 f^2} h_c^4 + \frac{7 c_8}{2 f^4}h_c^6 ~,\label{eq:masses} \\
m_\chi^2(h_c) &=& -\mu^2 +  \lambda\, h_c^2 + \frac{3 c_6}{4 f^2} h_c^4 + \frac{c_8}{2 f^4}h_c^6 ~,\label{eq:masses1} \\
m_t^2(h_c)&=&\frac{y_{t}^2}{2} h_c^2~,\quad m_W^2(h_c)=\frac{g^2}{4} h_c^2 ~,\quad m_Z^2(h_c)=\frac{g^2 +g'^2}{4} h_c^2 ~.\label{eq:masses2} 
\end{eqnarray}
%
 The explicit expression of the functions $J_b$ and $J_f$, with or without the hard thermal loop resummation, can be found \textit{e.g.} in Ref.~\cite{Quiros:1999jp,Laine:2016hma}.

Since our main results turn out to be quite insensitive to details, we can set the Yukawa, $SU(2)_L$ and $U(1)_Y$ gauge couplings at tree level by fixing $m_t(v)$, $m_W(v)$ and $m_Z(v)$ in Eq.~\eqref{eq:masses2} at 172, 80 and 91 GeV, respectively. For the mean-field estimates, in which zero-temperature one-loop corrections are neglected, we moreover constrain $\mu^2$ at tree level by requiring $V_{\rm tree}$ to have a minimum at $h_c=v$:
\begin{equation}
\label{eq:mu_mean8}
\mu^2=\lambda v^2 + \frac{3\, c_6}{4\, f^2} v^4 + \frac{c_8}{2\, f^4} v^6~.
\end{equation}
Similarly, to set $\lambda$, we require $\partial^2 V(\phi_c)/\partial h_c^2|_{h_c=v}=(125\,{\rm GeV})^2$, which implies
\begin{equation}
\label{eq:lam_mean8}
\lambda = -\frac{3\,c_6}{2\, f^2} v^2 - \frac{3\,c_8}{2\, f^4} v^4 + \frac{m_h^2}{2 v^2}~.
\end{equation} 
The remaining free parameters in $V_{1\ell}$ are therefore $c_6/f^2$ and $c_8/f^4$.

Notice that the EFT is a valid description of the theory only at energy scales much below $f$, therefore we do not address questions of the stability of the potential. Thus, we do not exclude \textit{a priori} all values of $c_8$ and $c_6$ leading to $V_{1\ell}$ unbounded from below; we only require 
\begin{equation}
\label{eq:stability}
V_{\rm tree}(v) < V_{\rm tree}(h_c) \quad \textrm{for any}~ h_c \in\, ]v,f]~.
\end{equation}
This in practice corresponds to imposing a lower bound on $c_8$ that varies with $f$. Such constraint is  $c_8\gtrsim -9$ for $f = 1$ TeV and $c_8\gtrsim -2$ for $f = 2$ TeV. For concreteness, we limit the plots hereafter to the first bound.

\begin{figure}[t]
\includegraphics[width=0.333\columnwidth]{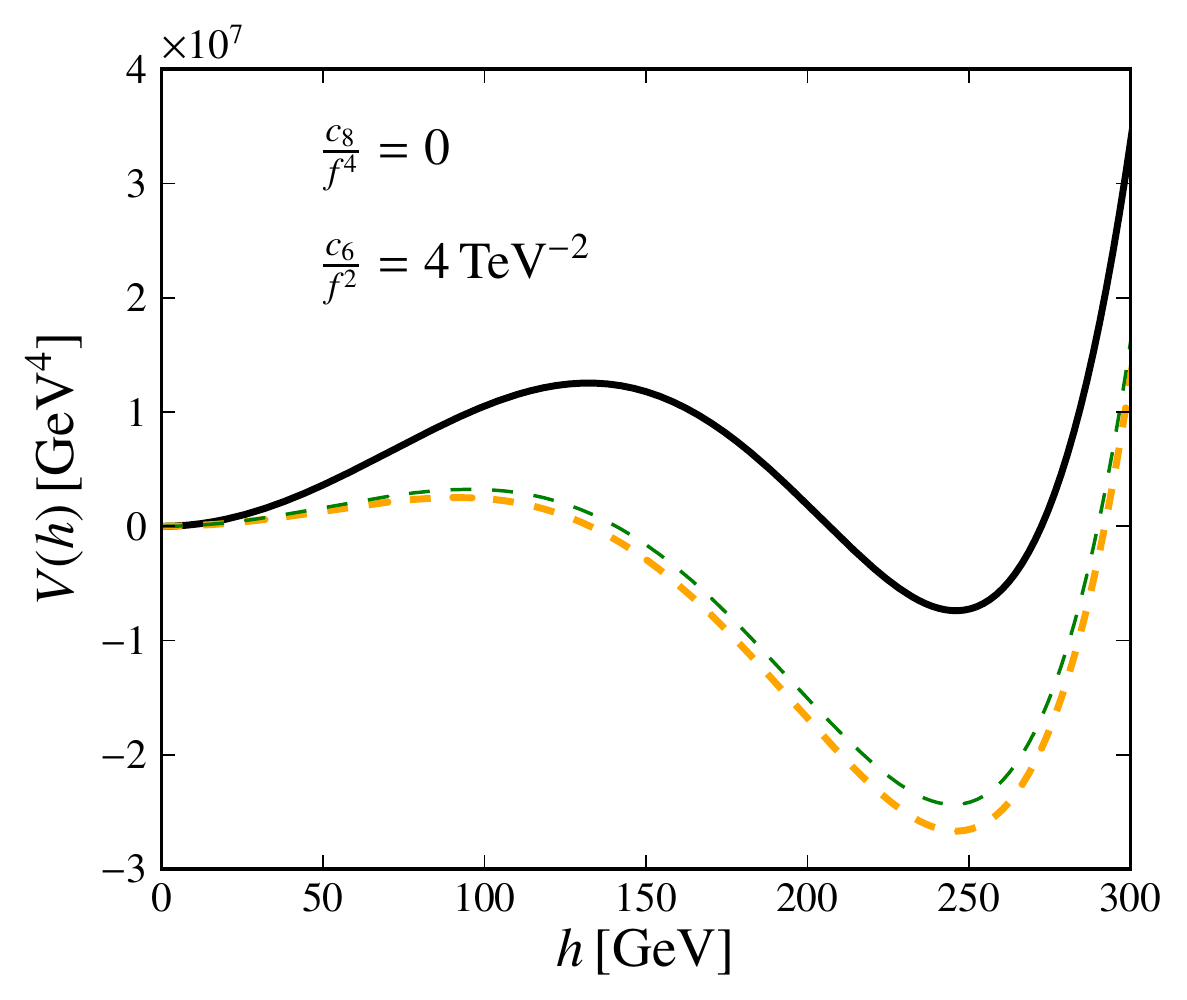}\hspace{-.1cm}
\includegraphics[width=0.333\columnwidth]{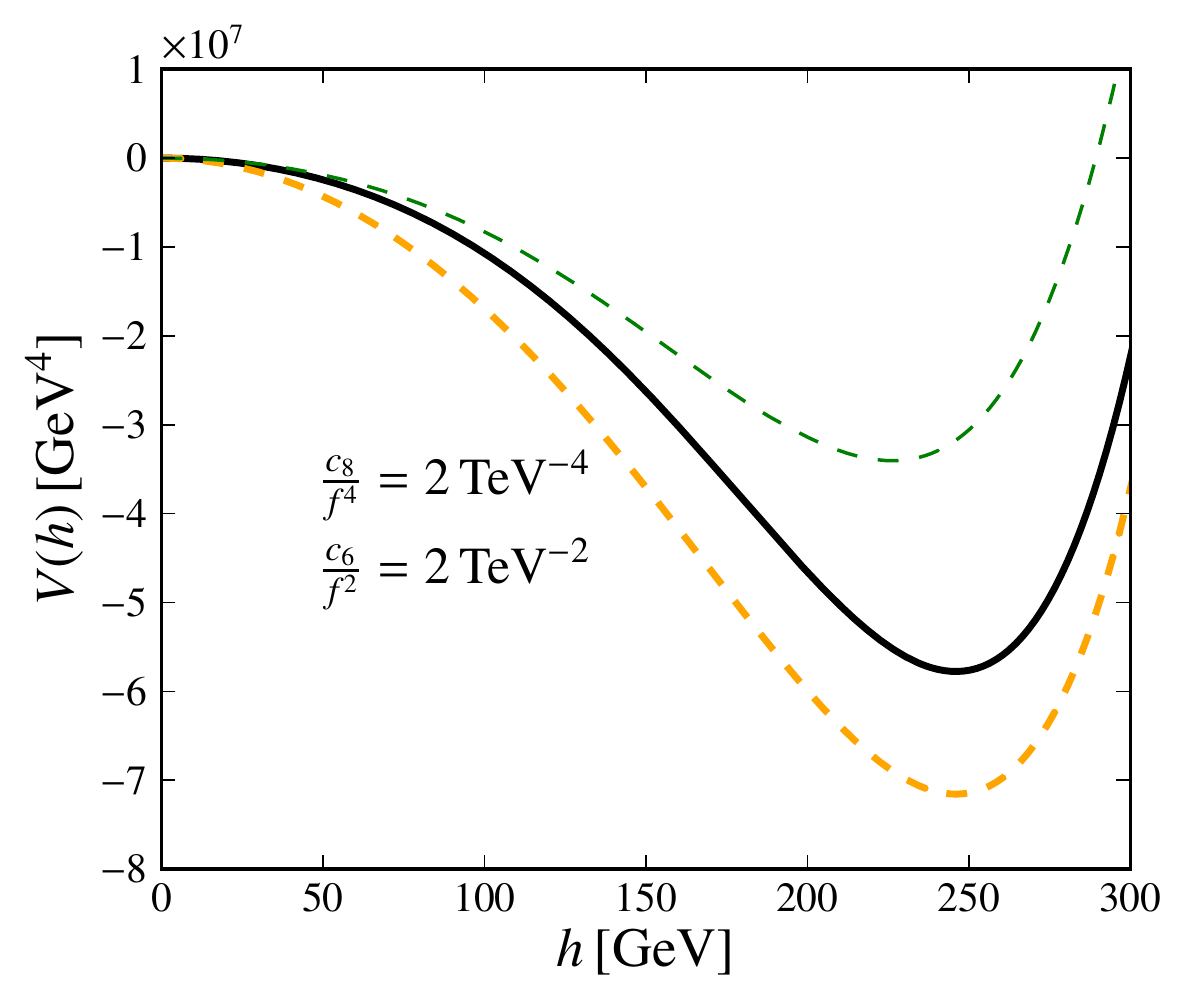}\hspace{-.1cm}
\includegraphics[width=0.333\columnwidth]{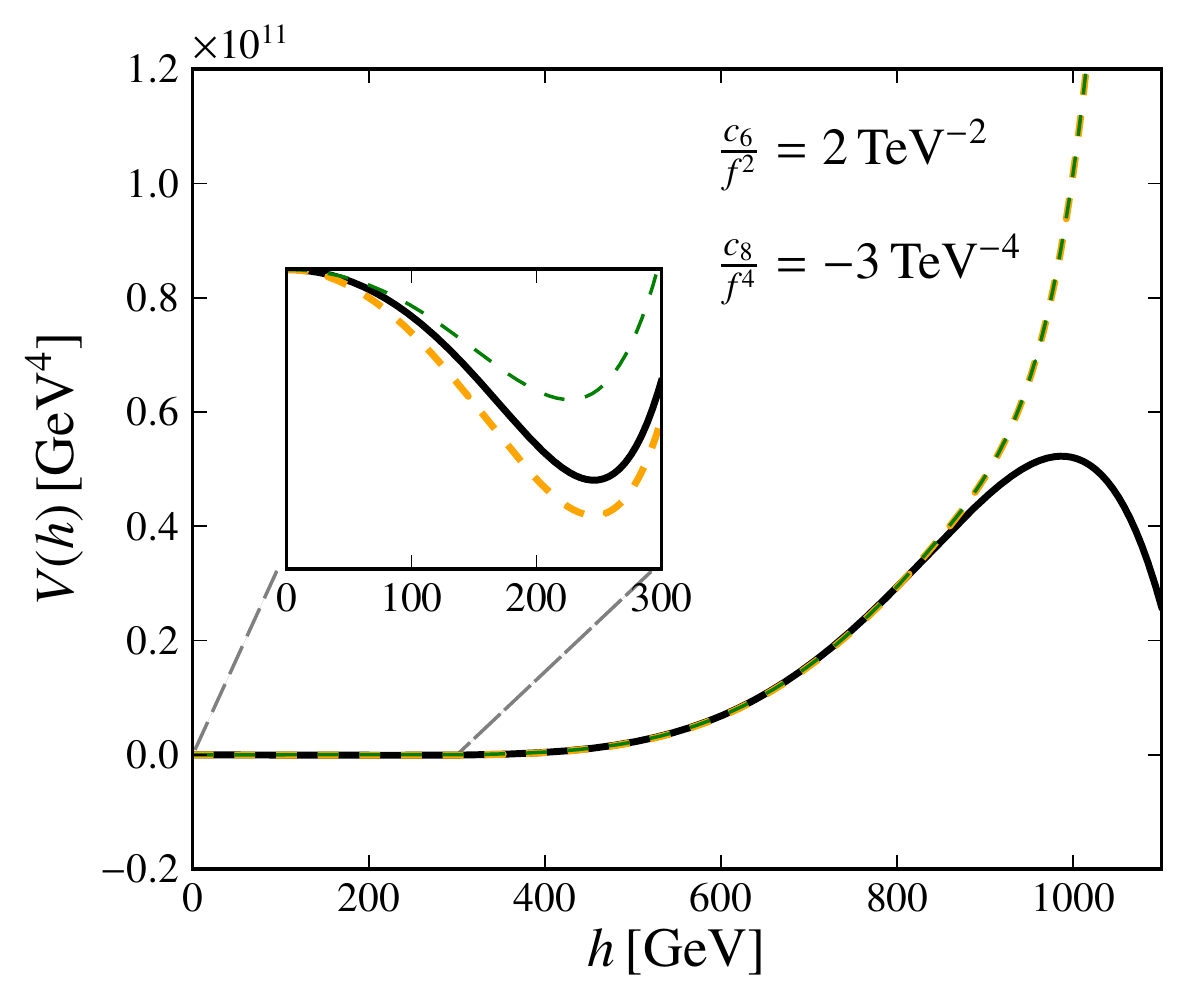}
\caption{\it The potentials $V_{\rm tree}$ (black solid curve), $V_{1\ell}$ at $T=0$ (orange dashed curve) and $V_{1\ell}$ at $T=T_x$ (green dashed curve) for the choices of $c_6/f^2$ and $c_8/f^4$ indicated in each panel. In the left panel, there exist two vacua already at zero temperature ($\mu^2\simeq -3100\,GeV^2$, $\lambda\simeq-0.23$, $T_x = T_c = 35\,GeV$). In the central panel, the existence of two vacua arises only at finite temperature  ($\mu^2=1900\,GeV^2$, $\lambda=-0.06$, $T_x=T_c=82\,GeV$). In the right panel, the potential is unbounded from below, but the instability scale is above the cutoff $f=1\,$TeV ($\mu^2=3000\,GeV^2$, $\lambda=-0.03$, $T_x = T_n=99\,GeV$). $T_c$ is the critical temperature obtained in the mean-field approximation.}\label{fig:V}
\end{figure}

Figure~\ref{fig:V} shows the typical classes of potentials that we consider: Cases where the potential has a tree level barrier between the minima (left panel), cases where such a barrier is only due to a finite temperature (one-loop) effect (central panel), and cases where the potential is unbounded from below but the instability arises at a scale larger than $f$ (right panel). See Ref.~\cite{Chung:2012vg} for phenomenological discussions of new physics models in each class.\\

\subsection{Mean-field estimates}
\label{subsec:mean}
From $V_{1\ell}$ it is straightforward to determine some quantities that roughly characterise the EWPT, namely $T_c$ and  $v_{T_c}/T_c$. The critical temperature, $T_c$, is the temperature at which the minima of the broken and unbroken phases are degenerate. It provides the upper bound on the temperature at which the EWPT really starts, $T_n$. 
The quantity $v_{T_c}/T_c$, with $v_{T_c}$ being the VEV of the Higgs in the EW broken phase at $T=T_c$, is linked to  
the strength of the EWPT. Indeed, due to the fact that $v_T/T$ typically decreases with increasing $T$, $v_{T_c}/T_c$ can be used as a lower bound on the actual value of $v_T/T$ during the EWPT (if the transition ever happens; see below). %

The potential $V_{1\ell}$ is easy to treat numerically, but for analytic insights on  $T_c$ and $v_{T_c}/T_c$, the mean-field approximation may be helpful. 
We then begin neglecting $\Delta_{1\ell,T=0} V(h_c)$. In  $\Delta_{1\ell,T\ne 0} V(h_c)$, we consider the high-temperature expansion of $J_b$ and $J_f$ and retain their leading terms, \textit{i.e.} $J_b(x)\to \pi^2 x/12$ and $J_f(x)\to -\pi^2 x/24$ in Eq.~\eqref{eq:VT}.
 The potential $V_{1\ell}$ now reduces to the form
\begin{equation}
\label{eq:V8app}
  V_{\rm mean}(\phi,T) = \frac{-\mu^2+a_T T^2}{2} h_c^2 +\frac{\lambda}{4} h_c^4 +\frac{c_6}{8 f^2} h_c^6+\frac{c_{8}}{16f^{4}} h_c^8~,  
\end{equation}
with $a_T = \frac{1}{16} \left(4\frac{m_{h}^{2}}{v^{2}}+3g^{2}+g'^{2}+4y_{t}^{2} -12 c_{6}\frac{v^2}{f^2} -12 c_8 \frac{v^4}{f^4} \right)$.

In Eq.~\eqref{eq:V8app} the thermal contribution can only raise the potential at $h_c\ne 0$. No transition from the symmetric to the broken phase is conceivable  if at zero temperature the EW breaking minimum is above the symmetric one. Hence, the condition $V_{\rm mean}(v,T=0)<V_{\rm mean}(0,T=0)$ has to be satisfied, which is equivalent to
\begin{equation}
\label{eq:c6_upper}
\frac{c_6}{f^2} < \frac{m_h^2}{v^4} - \frac{3 v^2}{2} \frac{c_8}{f^4}~.
\end{equation}
Saturating the inequality is not feasible. As previously mentioned, there must be a gap between $T_c$ and $T_n$, and the stronger the phase transition is the larger is the gap. For this reason, values of $c_6/f^2$ close to the upper bound in Eq.~\eqref{eq:c6_upper} are not acceptable since they lead to $T_c\to 0$ and $v_{T_c}/T_c \to \infty$. In this limit the EWPT would never happen within the lifetime of the Universe. Such values of $c_6/f^2$ are thus expected to be ruled out by more sophisticated estimates; see section \ref{subsec:numerics}. For the same reason, it is at large $c_6/f^2$ that, whenever the EWPT can really start, the parameter scenarios with the strongest  EWPTs arise. To appreciate the relevance of this effect, let us first evaluate the EWPT disregarding the issue.
%
%
%
%
%
%
%
%
 \begin{figure}[t]
    \begin{center} 
      \includegraphics[height=0.42\textwidth]{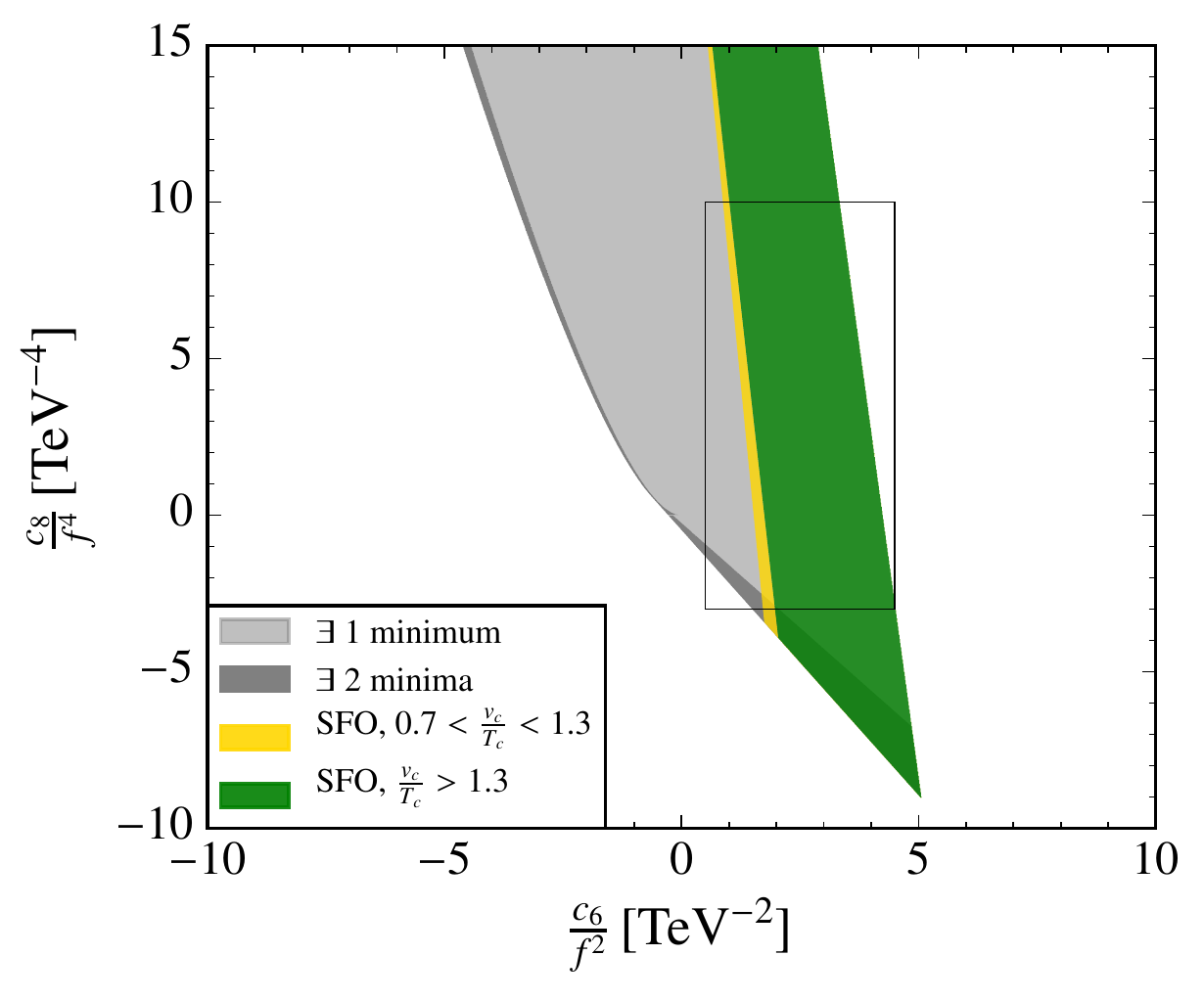}\hfill
      \includegraphics[height=0.42\textwidth]{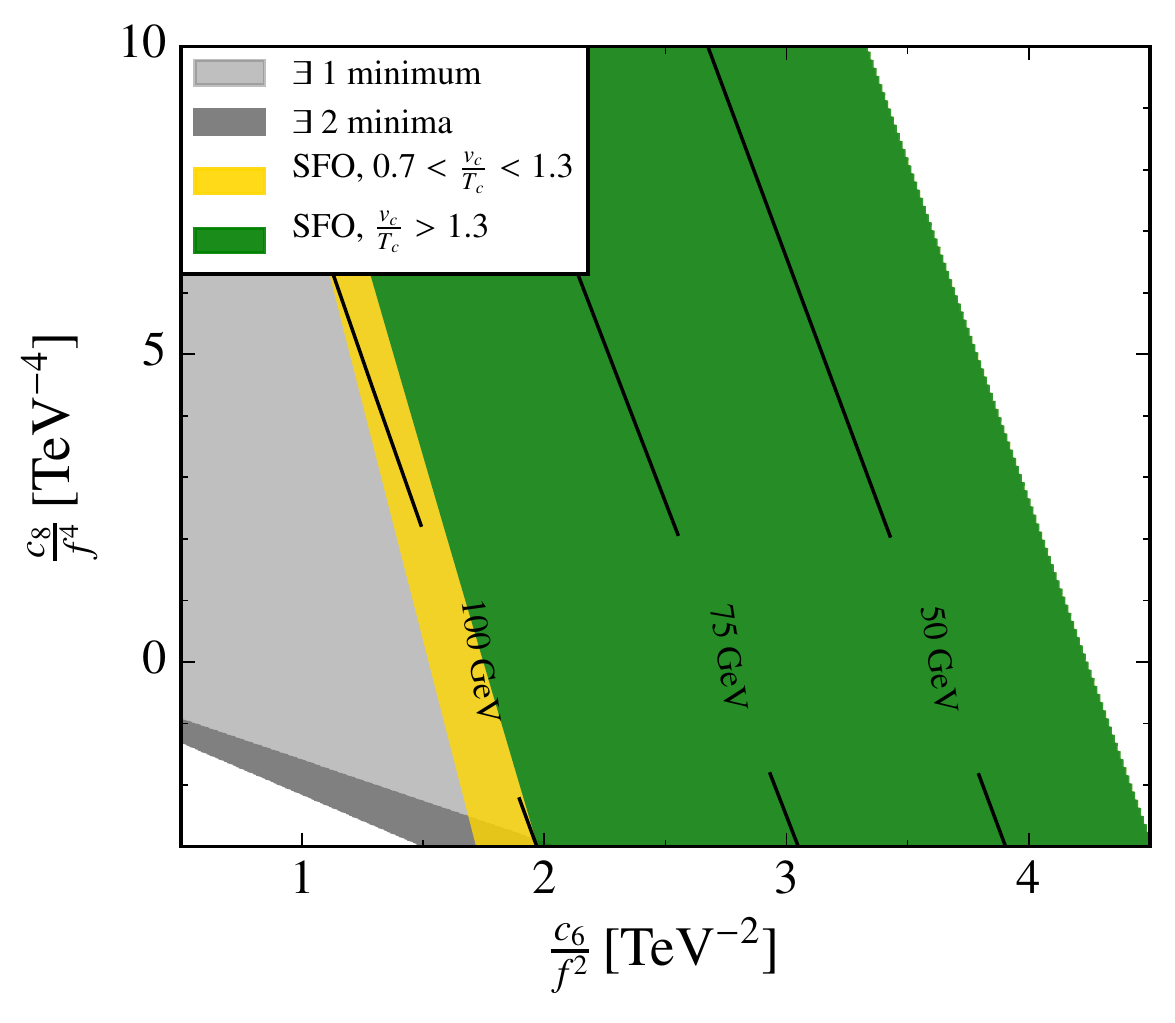}
    \end{center}
    \caption{\it $c_{6}/f^2-c_{8}/f^4$  of parameter space for a SFOEWPT in the mean-field approximation. Left) The filled region shows the allowed values for $c_{6}$ and $c_{8}$ such that at $T = 0$ the deepest minimum is at $v$. In the darker areas there is a second minimum above the one at $v$. For negative $c_{8}$, we cut off the potential at $1$ TeV and demand that $V(1 \text{TeV})>V(v)$ to ensure that the global minimum is at $v$. Superimposed are shades of yellow to green to show the strength of the phase transition, $v_{T_c}/T_{c}$, based on the critical temperature. Right) Zoomed version on the black rectangle of the left panel (note the different axis ranges). Lines of constant $T_c$ are depicted. %
    }
    \label{fig:c6c8}
  \end{figure}

We fix the values of $\mu$ and $\lambda$ as in Eqs.~\eqref{eq:mu_mean8} and \eqref{eq:lam_mean8}, and we require $c_6$ and $c_8$ to fulfil Eq.~\eqref{eq:stability}. Moreover, by definition, at $T=T_c$ the EWSB minimum is degenerate with the symmetric one. These properties lead to the following relations for $v_{T_c}$ and $T_c$:  
\begin{align}
  \begin{aligned}
    \label{eq:d-e:15}
    v_{T_c}^{2} & = \left[-\frac{2c_{6}}{3c_{8}}\pm 2\sqrt{\frac{c_{6}^{2}}{9c_{8}^{2}}-\frac{\lambda}{3 c_{8}}}\right]f^{2}~,\\
    T_c^2 &= \frac{\mu^2}{a_0}-\left[\frac{2 c_{6}^{3}}{27 c_{8}^{2}}-\frac{c_{6}\lambda}{3c_{8}}\mp 2 \sqrt{\left(\frac{c_{6}^{2}}{9c_{8}}-\frac{\lambda}{3}\right)^{3}\frac{1}{c_{8}}}\right]\frac{f^{2}}{a_0}~,
  \end{aligned}
\end{align}
The left panel of Fig.~\ref{fig:c6c8} summarises our mean-field-approximation results in the plane $c_6/f^2$--$c_8/f^4$.  %
To the right of the whole shaded area, Eq.~\eqref{eq:c6_upper} is violated. Therefore, along the right border, $T_c=0$ and $v_{T_c}/T_c=\infty$. On the left of it, the above conditions for a first order EWPT are not satisfied.  Below it, instead, Eq.~\eqref{eq:stability} is not satisfied for $f=1\,$TeV. (As previously explained, this border would move up or down by assuming different values of $f$.) The yellow and green regions mark the values of $c_6/f^2$ and $c_8/f^2$ leading to $0.7<v_{T_c}/T_c<1.3$ and  $v_{T_c}/T_c>1.3$, respectively. These regions are split into a darker and a lighter areas. For $c_8/f^2 <0$ the former shows where $V_{\rm tree}$ is unbounded from below but the instability is above the cutoff (\textit{cf}.~right panel in Fig.~\ref{fig:V}); in the latter, $V_{\rm tree}$ does not provide any sign of instability below the cutoff  (\textit{cf}.~left and central panels in Fig.~\ref{fig:V}). The same split is applied to the grey region where the EWPT is not strong. In the dark grey area with $c_8/f^2>0$, besides the global minimum at $h_c=v$, $V_{\rm tree}$ presents a further minimum at $h_c \in ]v,f[$. (For phenomenological implications of the latter see \textit{e.g.}~Ref.~\cite{Greenwood:2008qp}.) We do not further discuss this peculiar configuration since it does not appear in the region with a SFOEWT.  The right panel of Fig.~\ref{fig:c6c8} shows a zoom of the rectangle in the plot in the left panel. It also reports some contour curves for $T_c$.

\subsection{Numerical procedure}
\label{subsec:numerics}
The quantity $v_{T_c}/T_c$ is a good estimate of the strength of the EWPT only when the gap between $T_c$ and $T_n$ is small. Quantitatively, $T_n$ is defined as the temperature at which the probability for the nucleation of one single bubble (containing the broken phase) in a horizon volume is approximately $\sim 1$. For our scenario, the nucleation temperature can be considered in practice as the temperature $T_n$ such that $S_3[V_{1\ell}(h_c,T_n)]\simeq 140\,T_n$, with $S_3$ the action of the thermal decay from the false to the true vacuum of $V_{1\ell}$~\cite{Quiros:1999jp,Laine:2016hma}~\footnote{This assumes the Universe to be dominated by radiation during the EWPT.}. Analytically, $S_3$ can be calculated in the limit of thin or thick wall bubbles~\cite{Quiros:1999jp}, but in general we do not expect our bubble profiles to precisely fulfil any of these two limits. We thus determine $S_3$ numerically. For this scope, we use the code \texttt{CosmoTransition}~\cite{Wainwright:2011kj} in which, to be more accurate, we do not implement the potential in the mean-field approximation but as in Eq.~\eqref{eq:V1ell} with the hard-thermal loop resummations in $J_b$ and $J_f$ included~\footnote{We also modified the code to evaluate the $S_4$ bubble action. Within the numerical precision of the code, we did not find significant changes, at least in the resolution relevant for our plots.}. For this second and more precise study of the EWPT, for each value of $c_6/f^2$ and $c_8/f^4$ we determine numerically the values of $\mu$ and $\lambda$ for which $h_c = v$ and $m_h\sim 125$ GeV.

\begin{figure}[t]
\includegraphics[width=0.49\columnwidth]{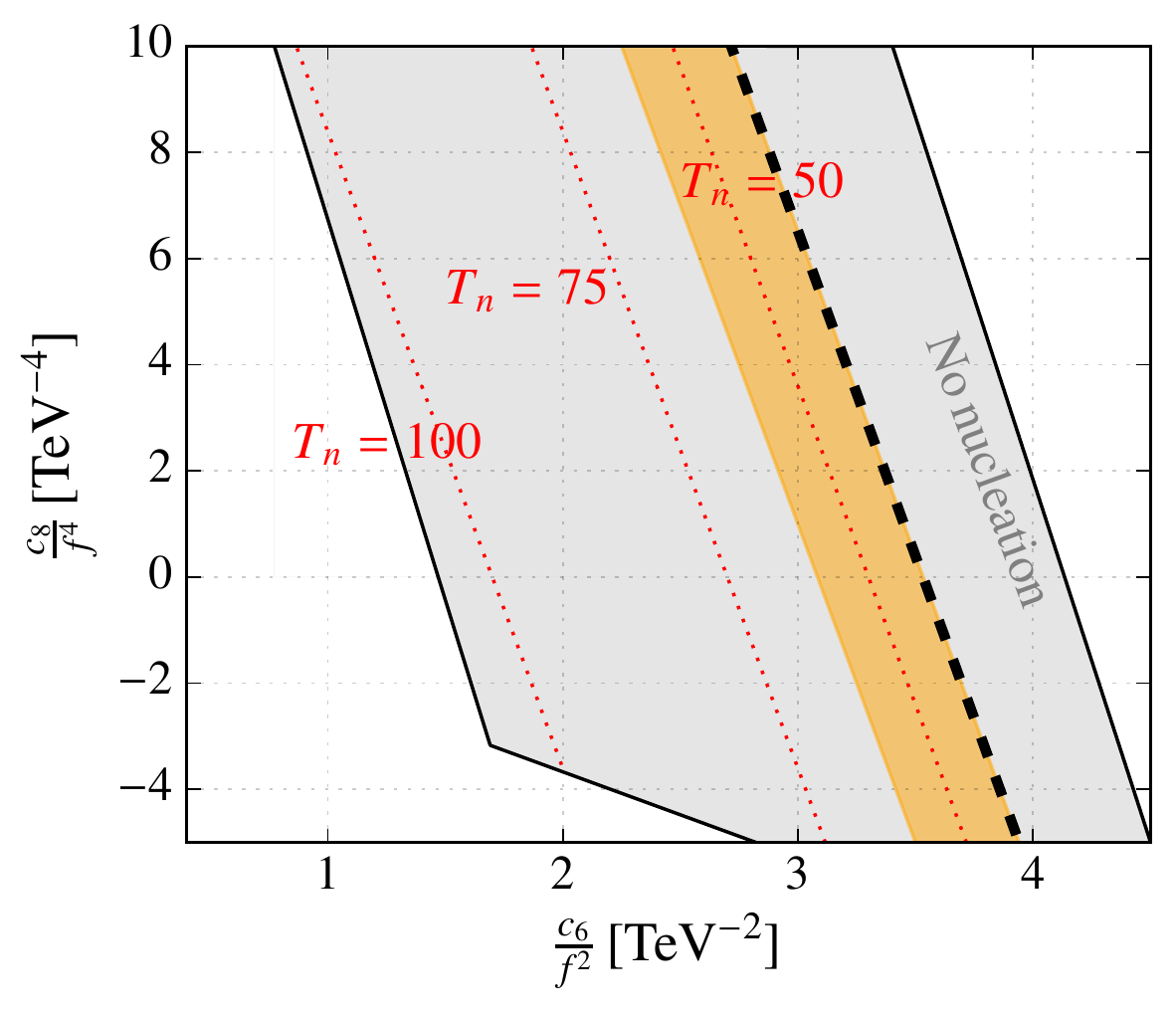}
\includegraphics[width=0.49\columnwidth]{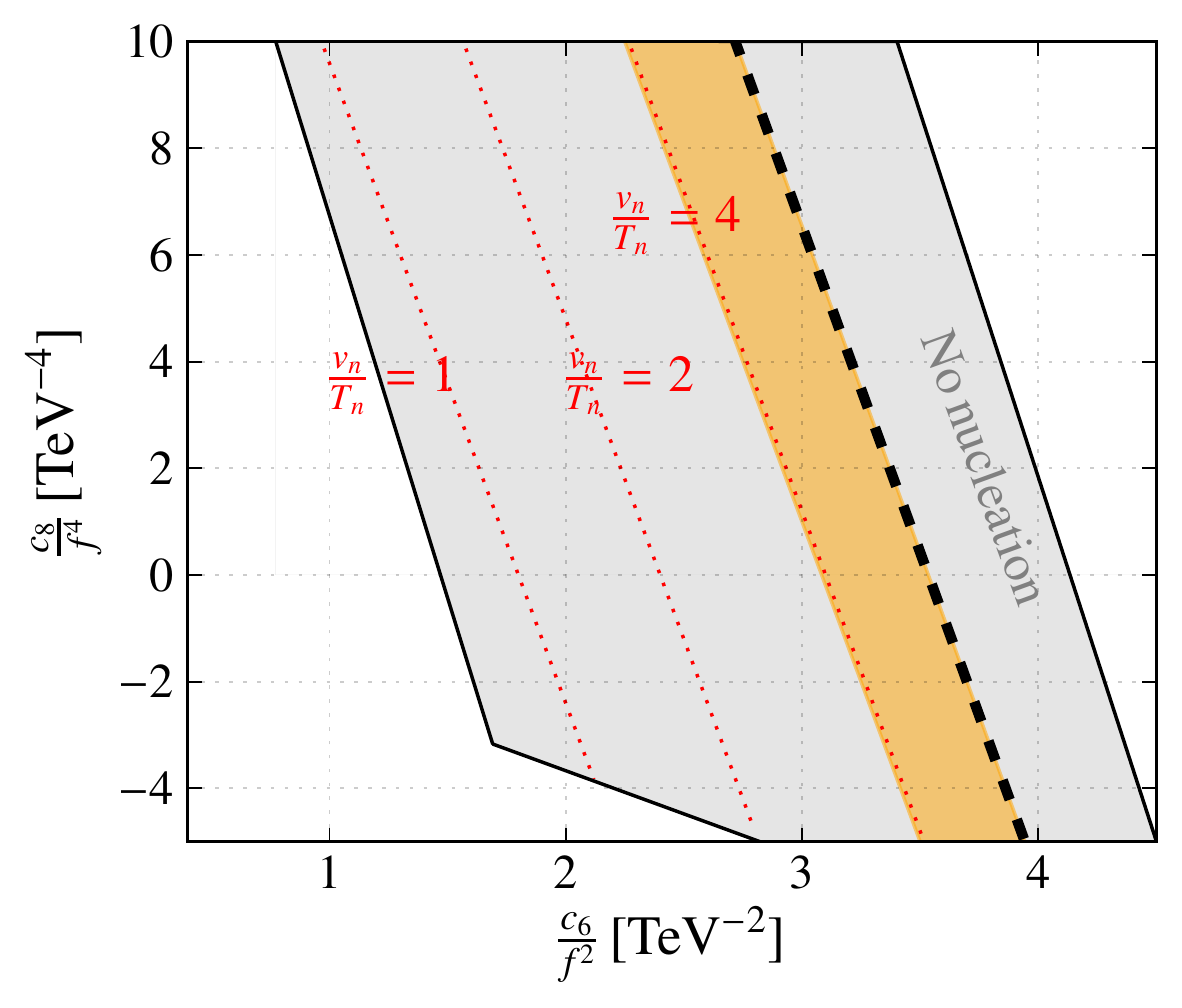}\\
\includegraphics[width=0.49\columnwidth]{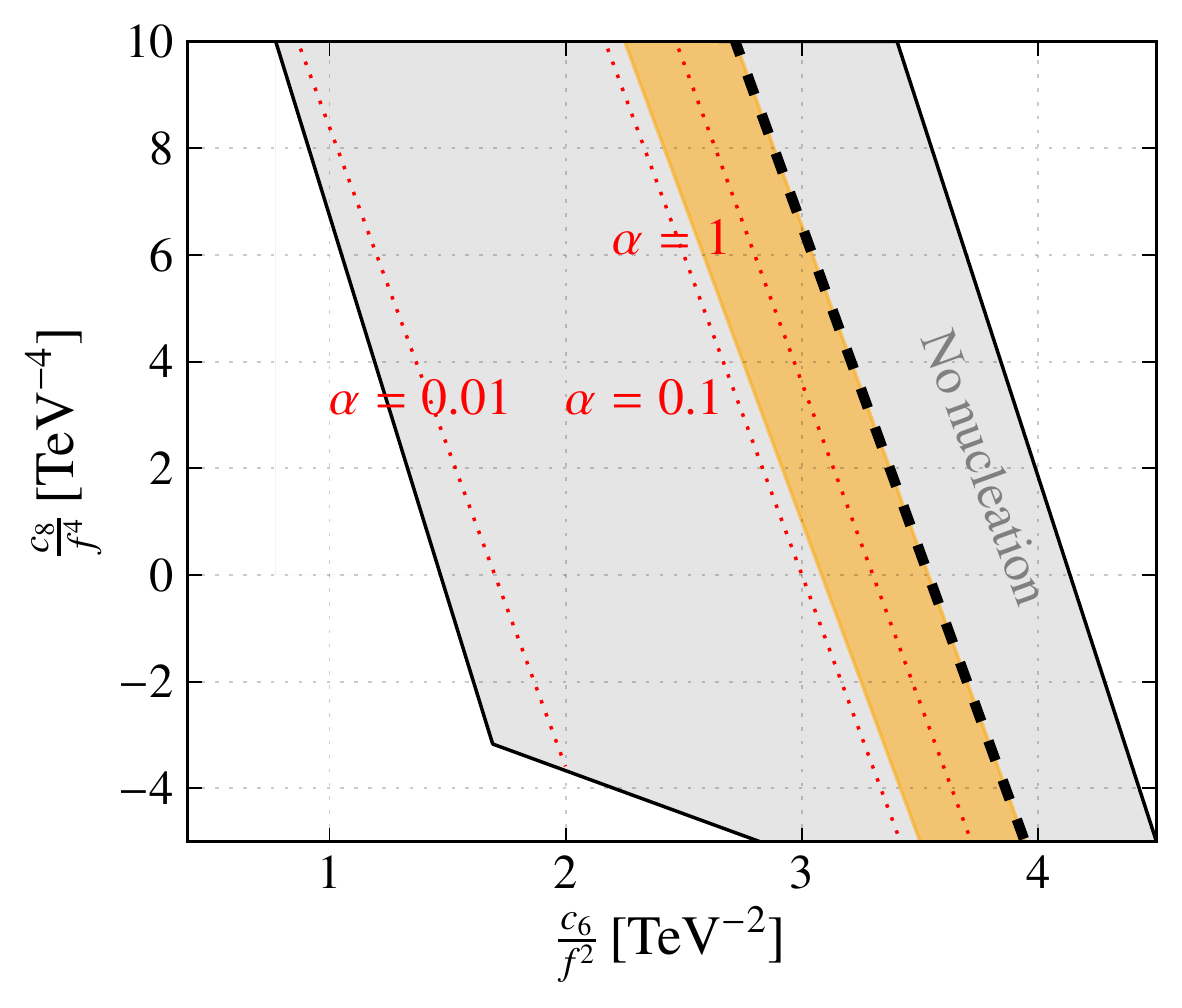}
\includegraphics[width=0.49\columnwidth]{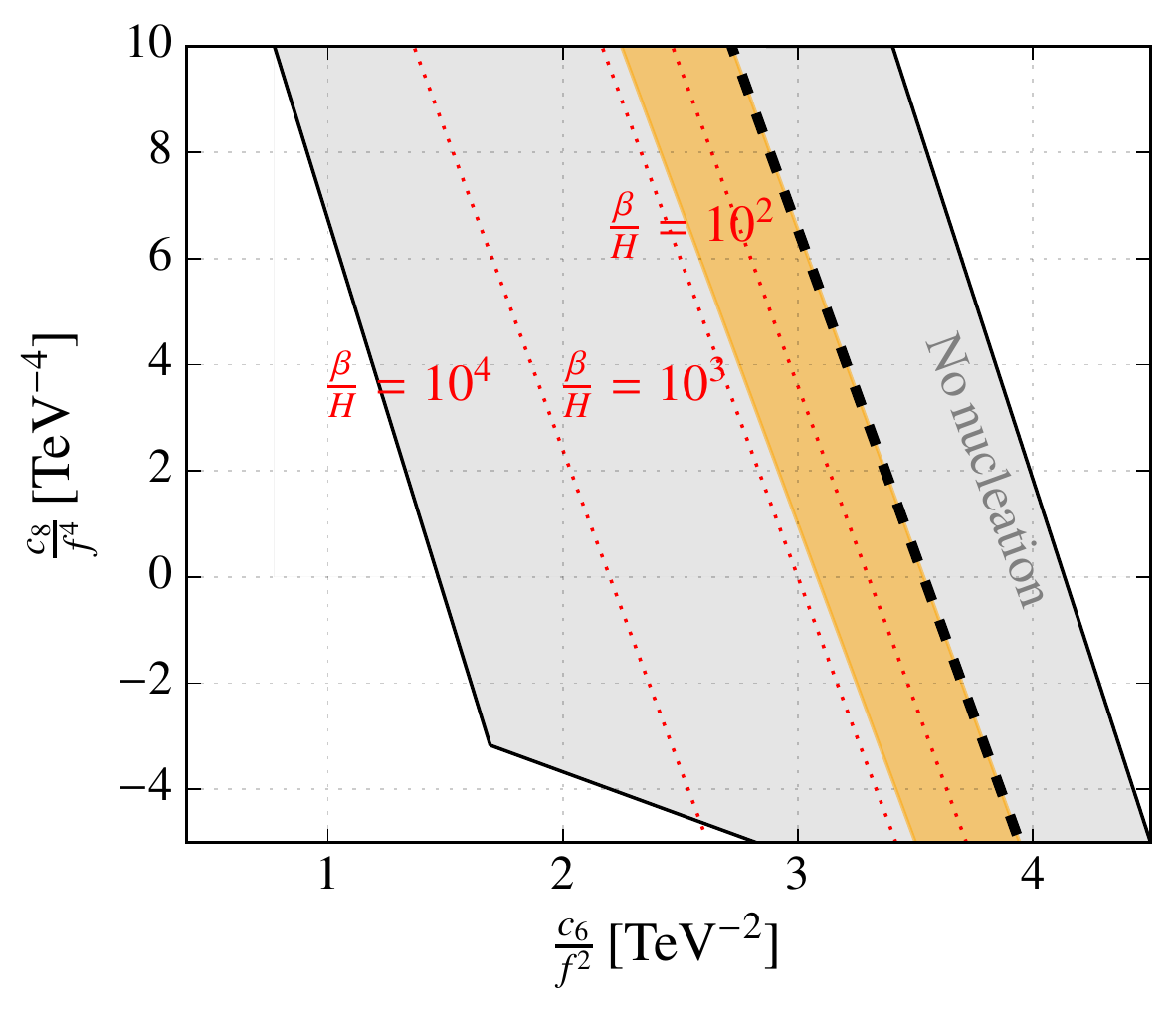}\\
\caption{\it Values of $T_n$ (top left), $v_{T_n}/T_n$ (top right), $\alpha$ (bottom left) and $\beta/H$ (bottom right) characterising the SFOEWPT in the plane $c_6/f^2$--$c_8/f^4$. The labels of $T_n$ and $T_c$ are in GeV units. On the right of the grey area the condition in Eq.~\eqref{eq:c6_upper} is violated. In the gray area to the right of the dashed line, the lifetime of the EW symmetric vacuum is longer that the age of the Universe, whereas on the left the transition results too weak for our purposes, \textit{i.e.} $v_{T_n}/T_n<0.7$. Below the grey area, the EW vacuum at zero temperature is not the global minimum at scales below the cutoff $f=1\,$TeV. In orange the parameter region LISA is sensitive to.}\label{fig:ewpt}
\end{figure}
The findings for $T_n$ and $v_{T_n}/T_n$ are respectively displayed in the top left and top right panels of Fig.~\ref{fig:ewpt}  (dotted lines). As expected, for values of $c_6/f^2$ nearby its upper limit (right border of the gray area; \textit{cf}.~Eq.~\eqref{eq:c6_upper}),  $S_3[V_{1\ell}(h_c,T)]/T$ is larger than 140 for any $T$, meaning that the EWPT never starts. This problem is avoided when $2c_6/f^2+ 3v^2c_8/f^4$ goes below the threshold of about $3.5$ (black, thick dashed line). Conceptually, at the threshold one obtains $T_n= 0$ and $v_{T_n}/T_n=\infty$. The strongest EWPTs and largest supercoolings (namely, the gaps between $T_n$ and $T_c$) are thus achieved just below this threshold. By departing from it (\textit{i.e.} by reducing $c_6/f^2$ at fixed $c_8/f^4$), the supercooling is reduced and, in turn, $v_{T_n}/T_n$ drops down. At some point, at about $c_6/f^2+ 3v^2c_8/(2 f^4) \approx 1.5$, the parameter  $v_{T_n}/T_n$ reaches 0.7, below which we do not draw any result. (We also omit the findings in the region where the EW vacuum instability is below the cutoff; see section~\ref{subsec:mean}.)
The values of $c_6/f^2$ and $c_8/f^4$ relevant for the present paper are therefore those within the gray and yellow regions on the left of the dashed thick line. 

\begin{figure}[t]
\includegraphics[width=0.49\columnwidth]{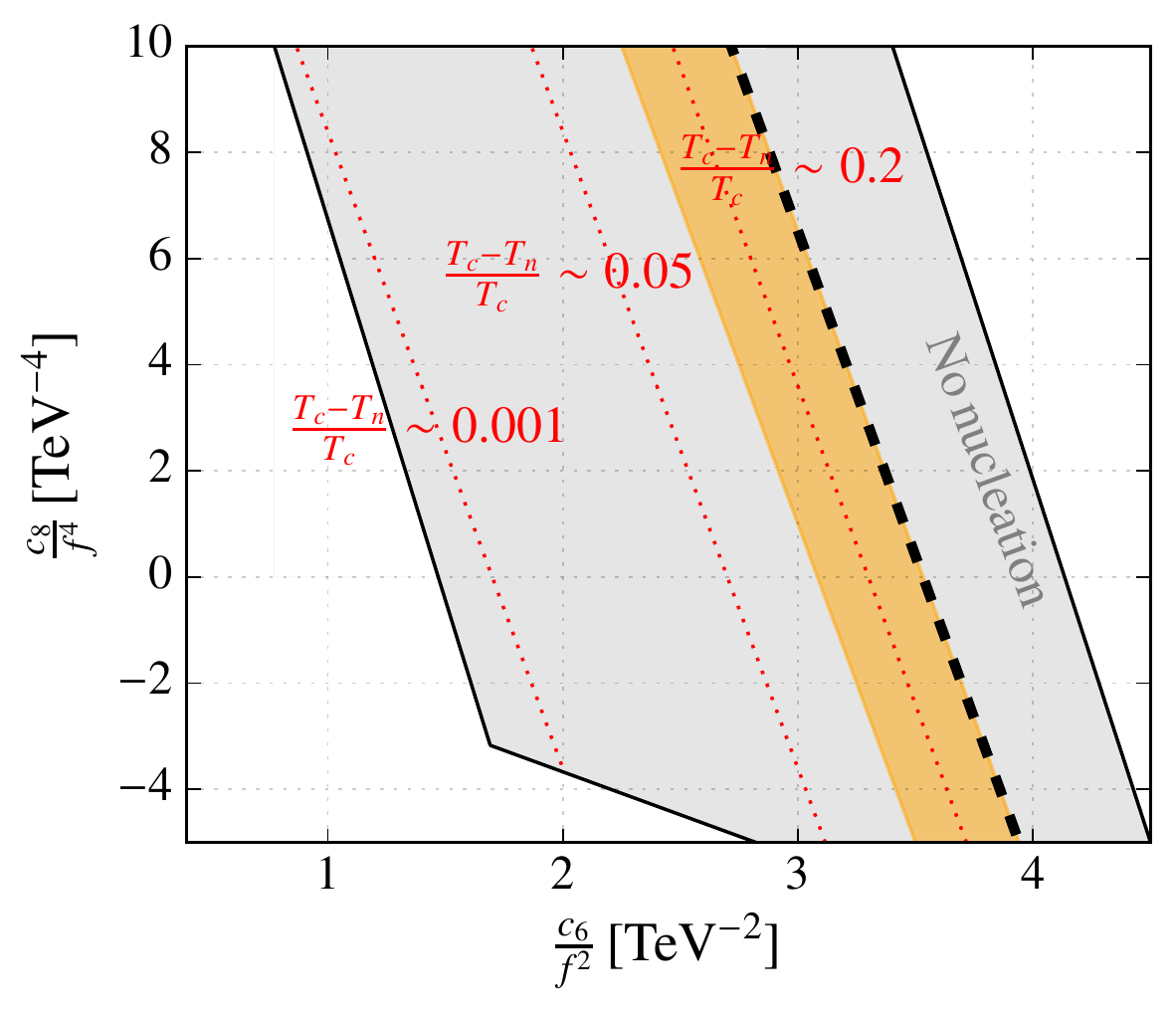}
\includegraphics[width=0.49\columnwidth]{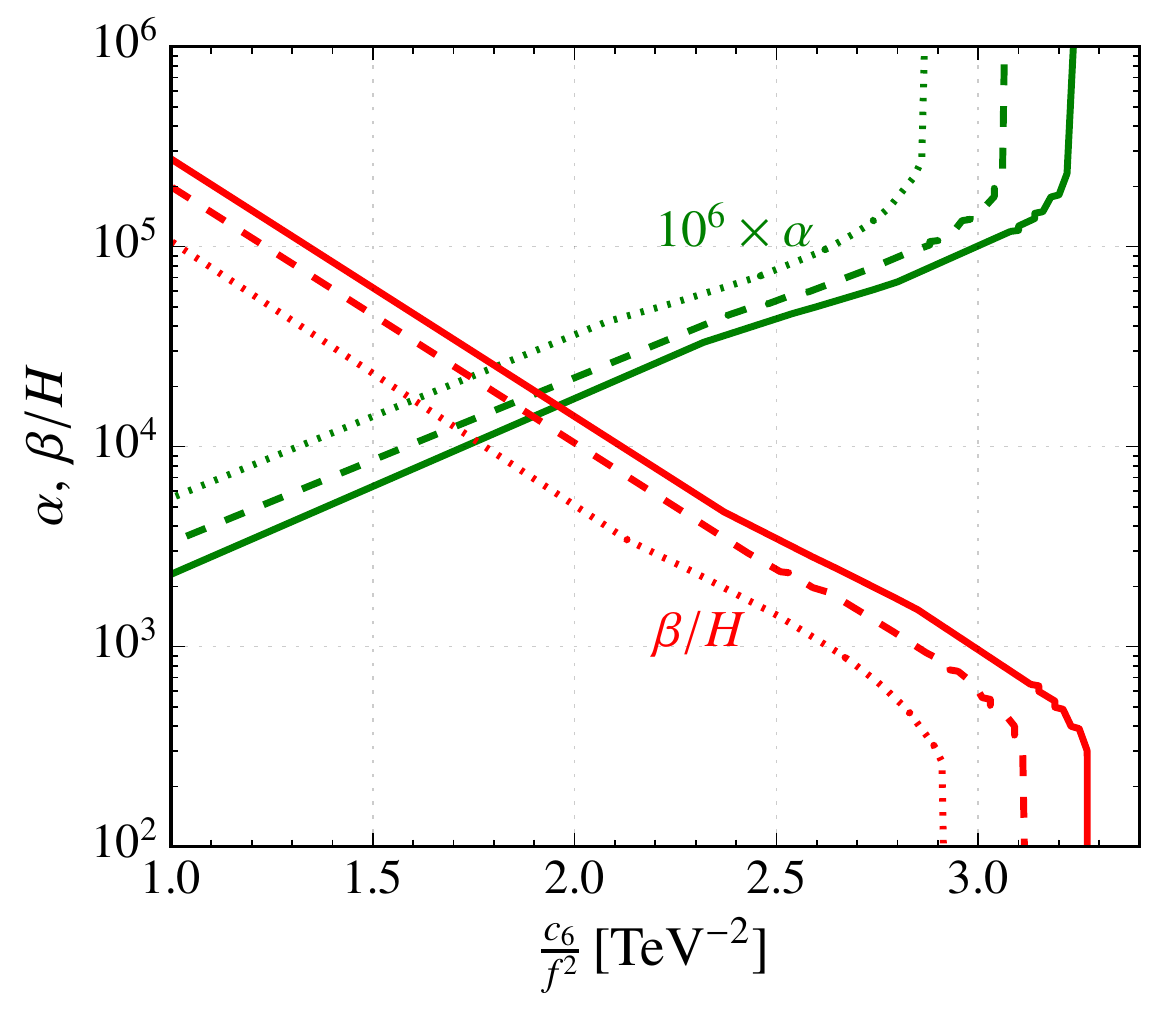}
\caption{\it Left panel) The values of $(T_c-T_n)/T_c$ in the plane $c_6/f^2$--$c_8/f^4$ (dotted lines). The rest stands as in Fig.~\ref{fig:ewpt}. Right panel) The values of $\beta/H$ and $10^6\times\alpha$ as a function of $c_6/f^2$ for $c_8/f^4=5\,{\rm TeV}^{-4}$ (dotted curves), $c_8/f^4=2\,{\rm TeV}^{-4}$ (dashed curves) and $c_8/f^4=0$ (solid curves).}\label{fig:ewpt2}
\end{figure}

The behaviour of $T_n$ and $T_c$ just described is also visible in the left panel of Fig.~\ref{fig:ewpt2}. As the figure highlights, for $v_{T_n}/T_n \gtrsim 4$ the discrepancy between $T_n$ (evaluated with the full potential $V_{1\ell}$ and hard-thermal loop resummation) and $T_c$ (evaluated in the mean-field approximation) is about 20\%, whereas negligible for $v_{T_n}/T_n\lesssim 1$. From this point of view, what prevents the use of $v_{T_c}/T_c\gtrsim 1$ in the mean-field approximation as a bound for EW baryogenesis (instead of $v_{T_n}/T_n\gtrsim 1$) is not the accuracy of the result but the presence of a sizeable region where the nucleation never occurs. 
Within the allowed $c_6/f^2$--$c_8/f^4$ parameter region, we also calculate the inverse duration time of the phase transition and the normalised latent heat. In our case we can approximate them, respectively, by $\beta/H= T_n \frac{d}{dT}\left( \frac{S_3}{T}\right)$ and $\alpha= \epsilon(T_n)/(35 T_n^4)$, where $\epsilon(T_n)$ is the latent heat at the temperature $T_n$. We determine them by means of \texttt{CosmoTransition}~\footnote{In order to obtain $\beta/H$ one has to modify the subroutine \texttt{transitionFinder.py}, as explained in Ref.~\cite{Chala:2016ykx}. Briefly, we determine $\beta/H$ by first finding the temperature $T_{240}$ at which $S_3[V_{1\ell}(h,T_{240})]/T_{240}=240$, and then we use the approximation $\beta/H\simeq T_n (240-140)/(T_{240}-T_n)$.}. Their dependencies on $c_6/f^2$ and  $c_8/f^4$ are presented in the bottom panels of Fig.~\ref{fig:ewpt}. The correlation between $T_n$, $v_{T_n}/T_n$, $\alpha$ and $\beta/H$ is evident. It is clear that all these quantities practically do not depend on $c_6/f^2$ and $c_8/f^4$ separately but only on $2c_6/f^2+ 3v^2c_8/f^4$. As expected, nearby the thick dashed line, where $T_n$ is small and $v_{T_n}/T_n$ is large, the EWPT exhibits small $\beta/H$ and  large $\alpha$, typical of large supercoolings. The values of $\alpha$ and $\beta/H$ that we obtain are more readable in Fig.~\ref{fig:ewpt2} (right panel) where their values are expressed as a function of $c_6/f^2$ for $c_8/f^4=5\,{\rm TeV}^{-4}$ (dotted curves), $c_8/f^4=2\,{\rm TeV}^{-4}$ (dashed curves) and $c_8/f^4=0$ (solid curves). In general, for $c_8= 0$, our results are in very good agreement with those of Ref.~\cite{Grojean:2004xa,Delaunay:2007wb}.

A further quantity useful to characterise the EWPT is $v_w$, the velocity at which the bubbles containing the broken phase expand into the EW symmetric phase. This speed results from the balance between the pressure difference between the two phases and the friction of the plasma on the bubbles.  
In general, the determination of $v_w$ is subtle~\cite{Bodeker:2004ws, Huber:2013kj, Konstandin:2014zta, Leitao:2014pda}. Fortunately, for our aim, it is relevant to know $v_w$ only when $v_{T_n}/T_n\gtrsim 4$; see below. In such a regime, on one side one expects $v\gtrsim 0.9$~\cite{Huber:2013kj}, on the other side $v_w$ cannot reach the speed of light, even asymptotically~\cite{Bodeker:2009qy,Bodeker:2017cim}. Due to this tiny window, it seems acceptable to take  $v_w=0.95$, for which we can straightforwardly adopt some results of the gravitational wave literature.

A SFOEWPT sources a gravitational wave stochastic background. Its power spectrum depends on $v_w$, $T_n$, $\beta/H$ and $\alpha$~\cite{Caprini:2015zlo}. If the amplitude of the signal is strong enough, the LISA experiment will be able to detect it towards the end of the LHC~\cite{Audley:2017drz}. Figure~4 in Ref.~\cite{Caprini:2015zlo} shows the values of $\beta/H$ and $\alpha$ that LISA can probe when $v_w\simeq 0.95$. We use this figure to forecast the capabilities of LISA for constraining the EFT we are working with~\footnote{The LISA design approved by ESA has a sensitivity that is quite similar to that dubbed ``C1`` in Fig.~4 of Ref.~\cite{Caprini:2015zlo}.  For our analysis we then use the ``C1'' sensitivity region of that figure. Moreover, as \textit{a posteriori} it turns out that LISA can probe our region when $T_n\lesssim 50\,$GeV, we use the result with $T_n=50\,$GeV of Ref.~\cite{Caprini:2015zlo}.}. The region that can be tested is marked in yellow in Figs.~\ref{fig:ewpt} and \ref{fig:ewpt2}.

\subsection{Interplay between gravitational wave signatures and Higgs-self coupling measurements}\label{sec:interplay}
From the bottom-up perspective we have adopted so far, the only collider implications of the operators $\mathcal{O}_6$ and $\mathcal{O}_8$ are changes in the rates of double- and triple-Higgs production. These are related to the modified Higgs couplings. Neglecting radiative corrections, the latter are given by
\begin{equation}
  \label{eq:h3h4}
  \frac{\lambda_{3}}{\lambda_{3,\text{SM}}} = 1+ \frac{v^{2}}{m_{h}^{2}}\left(2 c_{6} \frac{v^{2}}{f^{2}} +4c_{8} \frac{v^{4}}{f^{4}}\right)~, \quad
  \frac{\lambda_{4}}{\lambda_{4,\text{SM}}} = 1 + 4 \frac{v^{2}}{m_{h}^{2}} \left(3 c_{6} \frac{v^{2}}{f^{2}}+8c_{8}\frac{v^{4}}{f^{4}}\right) ,
\end{equation}
\begin{table}[t]
\begin{center}
\begin{tabular}{cc|cccc}
  $\frac{c_6}{f^2}\,[\text{TeV}^{-2}]$ & $\frac{c_8}{f^{4}}\,[\text{TeV}^{-4}]$ & $\lambda_3^{(0)}/ \lambda_{3,\text{SM}}$ & $\lambda_4^{(0)}/ \lambda_{4,\text{SM}}$ & $\lambda_3/ \lambda_{3,\text{SM}}$ & $\lambda_4/ \lambda_{4,\text{SM}}$\\
  \hline
  0 &  0 & 1    & 1    & 0.91 & 0.56 \\
  2 & -2 & 1.82 & 5.72 & 1.68 & 5.02 \\
  2 &  0 & 1.94 & 6.63 & 1.77 & 5.81 \\
  2 &  5 & 2.22 & 8.89 & 2.01 & 7.79 \\
  4 & -2 & 2.76 &11.34 & 2.53 &10.48 \\
  4 &  0 & 2.88 &12.25 & 2.63 &11.32 \\
  4 &  5 & 3.16 &14.52 & 2.87 &13.44 \\
\end{tabular}
\end{center}
\caption{\it Comparison between tree (denoted by the superscript `` $^{(0)}$") and loop level values of $\lambda_3$ and $\lambda_4$ with respect to their SM, tree-level values.}\label{tab:self}
\end{table}
The corresponding numbers at one loop, obtained numerically for several values of $c_6/f^2$ and $c_8/f^8$ are also shown in Tab.~\ref{tab:self}.

\begin{figure}[t]
    \begin{center} %
      \includegraphics[height=0.4\textwidth]{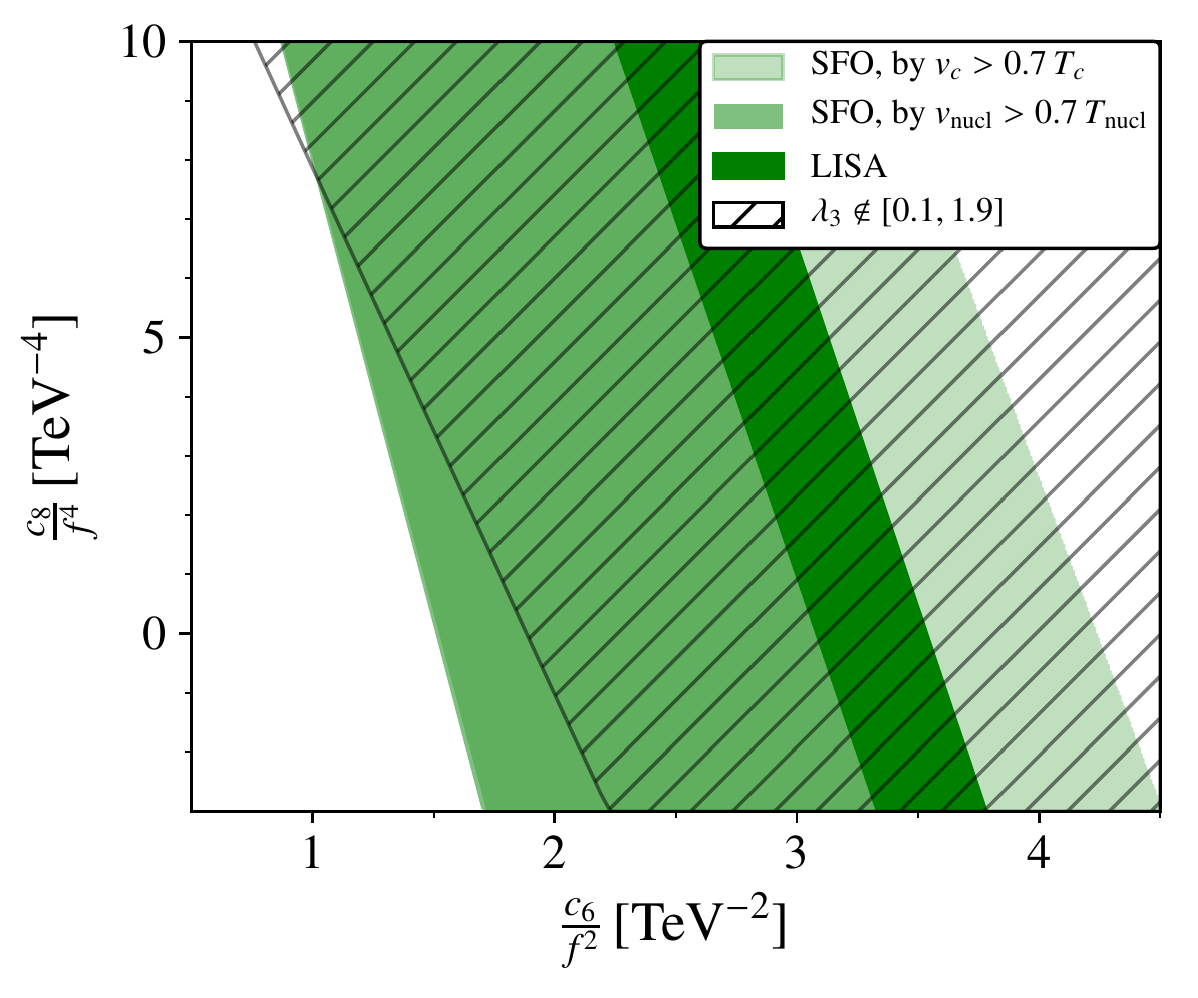}\hfill
            \includegraphics[height=0.4\textwidth]{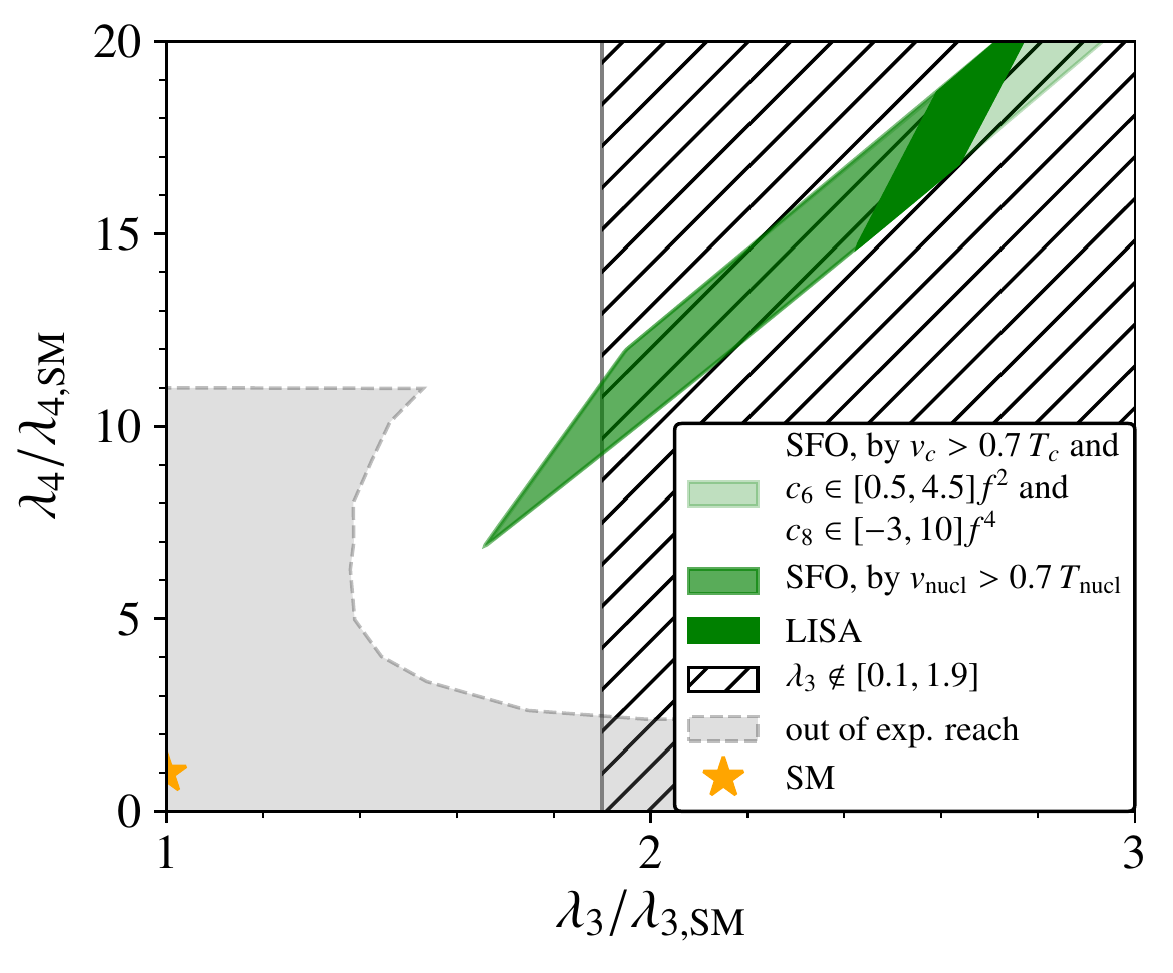}
    \end{center}
    \caption{\it Left panel) Region of Fig.~\ref{fig:c6c8} where the SFOEWPT is achieved accordingly to the criterion $v_{T_n}\gtrsim T_n$ instead of $v_{T_c}\gtrsim T_c$.   The reaches of FCC-ee~\cite{DiVita:2017vrr} and LISA~\cite{Caprini:2015zlo} are also displayed. Right panel) Allowed region from the left panel translated to the $\lambda_{3}/\lambda_{3,\text{SM}}$--$\lambda_{4}/\lambda_{4, \text{SM}}$ plane together with the future experimental sensitivities~\cite{Papaefstathiou:2015paa}. }
    \label{fig:l3l4}
  \end{figure}

These couplings have not been experimentally constrained yet. However, departures on the Higgs trilinear coupling beyond the range $[-0.7, 7.1]$ will be accessible at the 95\% C.L. in the HL-LHC run~\cite{DiVita:2017eyz,DiVita:2017vrr,Kim:2018uty}. Moreover, values outside the interval $[0.1, 1.9]$~\cite{DiVita:2017vrr} can be probed in a future FCC-ee facility~\cite{Gomez-Ceballos:2013zzn}. Likewise, searches for double-Higgs and triple-Higgs production at future hadron colliders might also constrain $\lambda_4$~\cite{Papaefstathiou:2015paa}. The reach of the different facilities is shown in the left panel of Fig.~\ref{fig:l3l4} as a function of $c_6/f^2, c_8/f^4$. In the right panel, this information is depicted in the plane $\lambda_{3}/\lambda_{3,\text{SM}}$--$\lambda_{4}/\lambda_{4, \text{SM}}$. The grey area in the latter shows the non-accessible region of a $100$ TeV $pp$ collider, taken from Ref.~\cite{Papaefstathiou:2015paa} (the reference cuts at $\lambda_{4}/ \lambda_{4,\text{SM}}=11$, and so do we). As we already mentioned, the region of the SFOEWPT identified by the nucleation temperature is a subset of the region found by the mean-field approximation. The region of parameter space that LISA is sensitive to is a subset of the former.\\  
With LISA starting to take data in the early 2030's, a sensible part of the parameter space where the SFOEWPT takes place would be first probed by LISA. Almost the complete parameter space would be tested at a future FCC-ee. A future hadron collider with $30~\text{ab}^{-1}$~\cite{Papaefstathiou:2015paa} could be fully conclusive.

\section{Matching of concrete models}
\label{sec:weakly}
The operators $\mathcal{O}_6$ and $\mathcal{O}_8$ are most commonly induced by new heavy scalars. These fields, however, generate normally other operators already at dimension six. Our aim here is to single out the properties of those UV completions that generate only $\mathcal{O}_6$ and $\mathcal{O}_8$ and  are allowed by current data. Let us parameterize the
effective Lagrangian after integrating out the new degrees of freedom as
\begin{equation}\label{eq:lag}
 L = L_{\text{SM}} + \sum_i \frac{c_i}{f^{4-d_i}}\mathcal{O}_{i}~,
\end{equation}
where $L_{\text{SM}}$ stands for the SM Lagrangian, $c_i/f^{4-d_i}$ represents the coefficient of 
the corresponding operator $\mathcal O_i$ and $f$ is the typical new-physics scale. The couplings are all expected to scale as $c_i \sim \tilde{g}^{2}\times \tilde{g}^{2n}/(4\pi)^{2n}$, with $\tilde{g}$ some weak coupling, $n$ the perturbative order at which $\mathcal O_i$ is generated, and $d_i$ the canonical dimension of $\mathcal O_i$. This means that $c_6$ can be $\mathcal{O}(1)$ TeV$^{-2}$ as required by the SFOEWPT only if the operator $\mathcal{O}_6$ is induced at tree level. Additionally, other operators with couplings of similar size will be generated.
Among these, we have, in a Warsaw-like basis~\cite{Grzadkowski:2010es}, the following ones~\cite{deBlas:2014mba}~\footnote{Note that $\mathcal O_6$ cannot be originated from integrating out at tree level new fermions~\cite{Jiang:2016czg,deBlas:2014mba,deBlas:2017xtg}. We also stress that the operator basis in Eq.~\eqref{eq:operators} is converted into the proper Warsaw basis~\cite{Grzadkowski:2010es} by integrating by parts $\mathcal{O}_{d6}$.}:
\begin{equation}
\label{eq:operators}
\mathcal O_{6} = (\phi^\dagger\phi)^3~,  
\quad \mathcal O_{d6} = \frac{1}{2}\partial_\mu(\phi^\dagger\phi)\partial^\mu(\phi^\dagger\phi)~, 
\quad \mathcal O_{\phi D} = (\phi^\dagger D_\mu \phi)((D^\mu\phi)^\dagger \phi)~.
\end{equation}
These typically appear together with further effective interactions. The same scalars generating $\mathcal O_{6},\mathcal O_{d6}$ and $\mathcal O_{\phi D}$ also induce, at the same order, the operators
\begin{equation}
\label{eq:O_psi}
\mathcal{O}_{\psi\phi} = y_\psi(\phi^\dagger\phi)(\overline{\psi}_L\phi \psi_R)~,
\end{equation}
with $y_\psi$ the Yukawa coupling of the SM fermions, here generically indicated as $\psi_L$ and $\psi_R$. These operators modifiy the Higgs-fermion interactions.
%

%
$\mathcal O_{d6}$ provides a contribution to the Higgs kinetic term. As a consequence, the Higgs couplings to fermions and gauge bosons are modified with respect to those of the SM by the factors~\footnote{In a complete dimension-six analysis, there are even more operators contributing to these factors, like $G_{\mu\nu}G^{\mu\nu}\phi^{\dagger}\phi$ and a similar operator for the photon. These can also be constrained by Higgs-couplings measurements and they do not contribute to the EWPT at tree level.}
\begin{equation}\label{eq:SMratio}
 \frac{g_{hff}}{g_{hff}^{\text{SM}}} = c, \quad \frac{g_{hVV}}{g_{hVV}^{\text{SM}}} = a, \quad \frac{g_{hgg}}{g_{hgg}^{\text{SM}}} = c, \quad \frac{g_{h\gamma\gamma}}{g_{h\gamma\gamma}^{\text{SM}}} = \frac{aI_\gamma + cJ_\gamma}{I_\gamma+J_\gamma}~,
\end{equation}
with
\begin{equation}
 a = 1 - c_{d6}\frac{v^2}{2f^2}, \quad c = 1 - c_{d6}\frac{v^2}{2f^2} + \mathcal{O}(c_{\psi\phi}, c_{\phi D})\frac{v^2}{f^2}~,
\end{equation}
and $I_\gamma \simeq -1.84$, $J_\gamma\simeq 8.32$. %
We can obtain
robust constraints on $c_{d6}$ from the present LHC measurements by marginalising the Run-2 constraints on $a$ over all possible values of $c$. One obtains~\cite{deBlas:2018tjm}
\begin{equation}
\label{eq:cd6_marginal}
c_{d6}\frac{v^2}{2 f^2}= -0.01\pm 0.06 \quad\rm{at~68\%~C.L.}~.
\end{equation}
A further improvement to $\pm 0.03$ is expected at the end of the HL run if no new physics is found \cite{deBlas:2018tjm}. %

We also note that neglecting $\mathcal{O}_{\phi D}$ can be justified at the matching scale, since  $c_{\phi D}(f)\approx 0$ can be naturally explained by means of UV symmetries. However, due to $\mathcal{O}_{d6}$, $c_{\phi D}$ runs between the renormalization scales $f$ and $v$~\cite{Alonso:2013hga}:
\begin{equation}
\label{eq:cd6.RGE}
c_{\phi D}(v) \simeq c_{\phi D}(f) + \frac{5}{24 \pi^2}g'^2 c_{d6}(f) \log{\frac{f}{v}}~.
\end{equation}
The present constraint on the coupling of $\mathcal{O}_{\phi D}$, namely~\cite{deBlas:2013gla}
\begin{equation}
-0.023 < c_{\phi D}/f^2~\text{TeV}^{2} < 0.006~, 
\end{equation}
provides an (indirect)  bound on $c_{d6}$.

Clearly, in view of the these bounds on $c_{d6}$ and $c_{\phi D}$, there will be little room for these couplings to be of the size of $c_6$, as suggested by power counting estimates in weakly-coupled scenarios. It is therefore crucial to understand whether there exist concrete UV scenarios that, at low energy, naturally generate a large hierarchy between $c_6$ and the other $c_i$ coefficients.
A hierarchy between different operator coefficients can also be generated (rather model-independently) with strongly-coupled UV-completions. We discuss the resulting picture in section ~\ref{sec:ewXL}.

\subsection{New scalars with weak isospin $I\leq 1$}
\label{sec:models}
In light of the above discussion, it is worth considering scenarios in which operators other than $\mathcal O_6$ are negligible. To this aim, let us first assume that the SM Higgs sector is extended only with new heavy scalars with isospin $I \leqslant 1$; see Ref.~\cite{Jiang:2016czg} for a related discussion. Concrete realisations and their signals at lepton colliders have been also discussed in Ref.~\cite{Cao:2017oez}. In the simplest case in which there is only one new field, $\varphi$, $\mathcal O_6$ is the only 
operator generated at tree level if and only if $\varphi$ is a colourless $SU(2)_L$ doublet with vanishing couplings to the fermions~\cite{deBlas:2014mba,DiLuzio:2017tfn}.
This scenario is then poorly motivated, because there is no symmetry that can remove \textit{only} the doublet couplings to fermionic currents, since a $\mathbb{Z}_2$ parity under which they are the only odd fields would make $c_6$ also vanish. Moreover, the new doublets appearing in the most common UV setups
do not exhibit this property.

On the other hand, one might argue that many motivated extensions of the SM Higgs sector involve several new fields. This is for instance the case of non-minimal composite Higgs models~\footnote{We note that composite Higgs models involve strongly-coupled dynamics and they are better described by the EW chiral Lagrangian; see section~\ref{sec:ewXL}. However, it has been shown that, in certain parameter space regions, the contribution of the extra scalars to the Higgs effective operators can overcome the contribution of the strong sector~\cite{Chala:2017sjk}.}. One particularly interesting example is the coset $SU(5)/SO(5)$~\cite{Vecchi:2015fma}, which admits a four-dimensional UV completion~\cite{Ferretti:2013kya}. The scalar sector consists of a hyperchargeless triplet, $\Xi_0$, a triplet with hypercharge $1$, $\Xi_1$, and a neutral singlet $\mathcal{S}$ on top of the Higgs doublet.
The effective operators we are interested in receive multiple contributions, namely
 \begin{align}
  \frac{c_{d6}}{f^2}  = \frac{1}{M^4}\bigg(\kappa_{\mathcal S}^2 - \kappa_{\Xi_0}^2 - 4|\kappa_{\Xi_1}|^2\bigg), \quad \frac{c_{\phi D}}{f^2} &= -\frac{2}{M^4}\bigg(\kappa_{\Xi_0}^2 - 2|\kappa_{\Xi_1}|^2\bigg)~,\\
  \frac{c_{\psi\phi}}{f^2} &= \frac{1}{M^4}\bigg(\kappa_{\Xi_0}^2 + 2|\kappa_{\Xi_1}|^2\bigg)~,
 \end{align}
and 
\begin{align}\nonumber
\frac{c_{6}}{f^2}&=\frac{\kappa_{\cal S}}{M^4}\left(-\lambda_{\cal S}\kappa_{\cal S}+\frac{\kappa_{{\cal S}^3}\kappa_{\cal S}^{2}}{M^2}-\lambda_{{\cal S}\Xi_0}\kappa_{\Xi_0}-4\operatorname{Re}\left[ \lambda_{{\cal S}\Xi_1}(\kappa_{\Xi_1})^*\right]+\frac{\kappa_{{\cal S} \Xi_0}\kappa_{\Xi_0}^2}{M^{2}} +\frac{2\kappa_{{\cal S}\Xi_1}|\kappa_{\Xi_1}|^2}{M^{2}}\right)\\\nonumber
&-\frac{ \kappa_{\Xi_0}^2}{M^4}\left(\lambda_{\Xi_0}-2\lambda\right)-\frac{\left|\kappa_{\Xi_1}\right|^2}{M^4}\left(2\lambda_{\Xi_1}-\sqrt{2}\tilde{\lambda}_{\Xi_1}-4\lambda\right)-\frac{2\sqrt{2}}{M^4}\operatorname{Re}\left[\lambda_{\Xi_1\Xi_0}(\kappa_{\Xi_1})^* \kappa_{\Xi_0}\right]\\
&-\frac{\sqrt{2}}{M^6 }\kappa_{\Xi_0\Xi_1}\kappa_{\Xi_0}|\kappa_{\Xi_1}|^2~,
\end{align}
where $M$ is the (assumed common) %
mass term of all new scalars, and the other couplings just parameterise the renormalizable interactions among themselves and 
the SM particles~\cite{deBlas:2014mba}.

It is interesting to show that not even in this case, which contains several scalars and many different couplings, can $\mathcal O_6$ be the only non-vanishing operator. Indeed, $c_{\phi D}$ only vanishes for $\kappa_{\Xi_0} = \sqrt{2}|\kappa_{\Xi_1}|$. This choice can in fact be enforced by an $SU(2)_L\times SU(2)_R$ symmetry, as in the Georgi-Machacek model~\cite{Georgi:1985nv}. It would yield
\begin{equation}
\frac{c_{d6}}{f^2} = \frac{1}{M^4}\bigg(\kappa_{\mathcal S}^2 - 6|\kappa_{\Xi_1}|^2\bigg)~,
\end{equation}
which could then be removed by enforcing $\kappa_{\mathcal S} = \sqrt{6}|\kappa_{\Xi_1}|$. As a result, it would turn out that $c_{\psi\phi}/f^{2} = 4|\kappa_{\Xi_1}|^2/M^4$, which vanishes if and only if $\kappa_{\Xi_1} = 0$. In such a case, however, $c_6$ is vanishing too.
%

%
Actually, we can go further and show that \textit{there is no weakly-coupled renormalizable extension of the Higgs sector containing singlets or triplets ---with non-vanishing couplings to the SM--- in which the effective operators produced after integrating out all new scalars at tree level modify only the scalar potential.} 

In order to prove this statement, let us work in the Warsaw basis and use the results of Ref.~\cite{deBlas:2014mba}. Let us also assume first that the extended Higgs sector contains (at least) one neutral singlet. This field generates a positive $c_{d6}$ that can be only cancelled by the contribution of a colourless triplet scalar. Indeed, any combination of colourless-triplet scalars, independently of the number of fields and their quantum numbers, gives a negative contribution to $c_{d6}$. 
This contribution is in fact the sum of all independent contributions~\cite{deBlas:2014mba}.

Colourless triplet scalars, on their side, also produce the operator $\mathcal O_{\psi\phi}$ with coefficient $c_{\psi\phi}\propto c_{d6}$. Therefore, it cannot be neglected if the triplet has to cancel the singlet contribution to $\mathcal O_{d6}$. The operator  $\mathcal O_{\psi\phi}$, in turn, cannot be cancelled by the singlet, which does not produce it at all at tree level. For this matter, at least one extra doublet is to be present, too. However, doublets produce also four-fermion operators like $\mathcal O_{le} = (\overline{l_L}\gamma_\mu l_L)(\overline{e_R}\gamma^\mu e_R)$. This is actually generated only by doublets, with negative sign for $l_L$ and $e_R$
of the same flavour. So, it cannot be removed at all by including other scalar fields. Instead, its coefficient must be explicitly forced to vanish. In such a case, however, the coupling $c_{\psi\phi}$ induced by the triplets would be strictly vanishing, and so all the linear interactions between the new physics and the SM, in contradiction with our hypothesis. Had we started considering the presence of at least one triplet, instead of one singlet, we would have arrived to exactly the same conclusion.

\subsection{New scalars with weak isospin $I> 1$}
Let us now consider the case $I > 1$. The only scalars that can couple in a renormalizable way to the SM sector are quadruplets with hypercharges $Y = 1/2, 3/2$. Interestingly, they contribute only to $\mathcal{O}_6$ when integrated out. These quadruplets can appear, for example, in Grand Unified Theories (GUT).

In GUT models, the SM fermions as well as the Higgs doublet are embedded in multiplets of a simple gauge group containing the SM $SU(3)_c\times SU(2)_L\times U(1)_Y$. 
Two main GUT gauge groups have been typically considered in the literature, namely $SU(5)$ and $SO(10)$ (and at a lesser extent, $E_6\supset SO(10) \supset SU(5)$). The minimal irreducible representations of the scalar fields that can lead to SM gauge uncoloured quadruplets are the $\mathbf{35}$ and the $\mathbf{70}$ in $SU(5)$~\cite{Slansky:1981yr,Feger:2012bs}. %

Obviously, such large-dimensional representations do not decompose only into quadruplets, but into many other states. An example is
\begin{equation}\label{eq:su5}
 \mathbf{35} = (\mathbf{1}, \mathbf{4})_{3/2} + (\mathbf{\overline{3}}, \mathbf{3})_{2/3} + (\mathbf{\overline{6}}, \mathbf{2})_{−1/6} + (\mathbf{\overline{10}}, \mathbf{1})_{−1}~,
\end{equation}
where the two numbers in parenthesis and the sub-index denote the dimension of the irreducible representation of $SU(3)_c$ and $SU(2)_L$ under which the corresponding field transforms and its hypercharge, respectively. Clearly, larger representations reduce to a larger number of exotic fields. Despite being unlikely, it is still possible that the effective operators generated by the coloured scalars are sub-leading with respect to the $\mathcal{O}_6$ induced by the quadruplet. This can happen it two cases: \textit{i)} If the coloured scalars are much heavier (which can be justified if a specific mechanism, similar to those advocated to solve the doublet-triplet splitting problem in SUSY GUT models~\cite{Witten:1981kv,Dvali:1993yf,Georgi:1981vf,Masiero:1982fe,Babu:2006nf,Babu:1993we}, is enforced); \textit{ii)} if all non-quadruplet fields have vanishing linear couplings to the SM at the renormalizable level. Surprisingly, this is the case for all extra fields in Eq.~\eqref{eq:su5} (although in principle they could couple, \textit{e.g.}, to dangerous flavour-violating currents via effective interactions).

Although the representation $\mathbf{35}$ does not include the Higgs boson, nor is required to break $SU(5)$ down to the SM gauge group (unlike \textit{e.g.}~the $\mathbf{24}$),  the aforementioned observations motivate further studies of a Higgs sector extended with quadruplets~\footnote{Larger representations, such as the mentioned $\mathbf{70}$, do contain a Higgs doublet, but also other fields with renormalizable interactions to the SM fermions. Moreover, smaller representations typically contain singlets and triplets (such as in the $\mathbf{15}$ and the $\mathbf{24}$, to name a few).}. There is a caveat, though. Despite being suppressed by higher powers of $1/M^2$, with $M$ the mass of these fields, dimension-eight operators can be also in conflict with current data. For example, for a quadruplet with $Y = 3/2$, the operator $(\phi^\dagger\phi)\mathcal{O}_{\phi D}$, which violates custodial symmetry at dimension eight, carries a coefficient of order $\sim c_6/M^4$. The rather low upper bound on $M\lesssim$ few hundred GeVs implied by the SFOEWPT is therefore in tension with the very well measured value of the $\rho$ parameter~\cite{Babu:2009aq,Ghosh:2016lnu}. Indeed, the experimental bound $\rho_{\textrm exp}=(1.00037\pm 0.00023)$~\cite{Patrignani:2016xqp} imposes $M \gg 1 $ TeV. A way out to this problem is considering a custodially symmetric quadruplet setup. We devote next section to this topic.

\subsubsection{A custodial quadruplet setup}
We start from the custodially symmetric Lagrangian of the $SU(2)$-quadruplet that was discussed in Ref.~\cite{Logan:2015xpa}. The potential is~\footnote{We use the same convention and notation for the generators as in Ref.~\cite{Logan:2015xpa}.}
\begin{align}
\begin{aligned}
  \label{eq:cus-qdrpt:1}
  \mathcal{L} &= \frac{1}{2} \langle (D_{\mu}\Theta)^{\dagger} D^{\mu}\Theta\rangle + \frac{1}{2} \langle (D_{\mu}\Phi)^{\dagger} D^{\mu}\Phi\rangle - \frac{\mu^{2}}{2} \langle\Phi^{\dagger}\Phi\rangle - \frac{\lambda}{4} \langle\Phi^{\dagger}\Phi\rangle^{2}\\
  & -\frac{\mu^{2}_{\Theta}}{2} \langle\Theta^{\dagger}\Theta\rangle- \frac{\lambda'}{4} \langle\Phi^{\dagger}\Phi\rangle \langle\Theta^{\dagger}\Theta\rangle- \widetilde{\lambda} \langle \Phi^{\dagger}T^{a}_{1/2}\Phi T^{b}_{1/2}\rangle \langle \Theta^{\dagger}T^{a}_{3/2}\Theta T^{b}_{3/2}\rangle\\
  & - \frac{2 \sqrt{2}}{3} \lambda_{\Theta} \langle \Phi^{\dagger} \hat{T}^{1,a}_{1/2} \Phi (\hat{T}^{1,b}_{1/2})^{\dagger}\rangle \langle \Phi^{\dagger} (\hat{T}^{1,a}_{3/2\, 1/2})^{\dagger} \Theta \hat{T}^{1,b}_{3/2\, 1/2} \rangle\\
    & - \frac{2 \sqrt{2}}{3} \lambda_{\Theta} \langle \Phi \hat{T}^{1,a}_{1/2} \Phi^{\dagger} (\hat{T}^{1,b}_{1/2})^{\dagger}\rangle \langle \Phi (\hat{T}^{1,a}_{3/2\, 1/2})^{\dagger} \Theta^{\dagger} \hat{T}^{1,b}_{3/2\, 1/2} \rangle \quad +\mathcal{O}(\Theta^{3},\Theta^{4}),
  \end{aligned}
\end{align}
where $\langle A \rangle$ is the trace of the matrix $A$, and
\begin{align}
\begin{aligned}
  \label{eq:cus-qdrpt:2}
  \Theta = \begin{pmatrix}\Theta_{3}^{*} & -\Theta_{1}^{-*}&\Theta_{1}^{++}&\Theta_{3}^{+++}\\-\Theta_{3}^{+*}&\Theta_{1}^{*}&\Theta_{1}^{+}&\Theta_{3}^{++}\\ \Theta_{3}^{++*}&-\Theta_{1}^{+*}&\Theta_{1}&\Theta_{3}^{+}\\ -\Theta_{3}^{+++*}&\Theta_{1}^{++*}&\Theta_{1}^{-}&\Theta_{3} \end{pmatrix}\equiv\begin{pmatrix}\widetilde{\Theta}_{3} & \widetilde{\Theta}_{1}&\Theta_{1}&\Theta_{3}\end{pmatrix} \quad \text{and} \quad \Phi = \begin{pmatrix}h_{0}^{*}&h^{+}\\ -h^{-}&h_{0} \end{pmatrix}\equiv\begin{pmatrix}\widetilde{\phi}&\phi\end{pmatrix}.
  \end{aligned}
\end{align}
In this notation, $\Theta_{3}$ has hypercharge $3/2$, $\Theta_{1}$ has hypercharge $1/2$, and $\phi$ is the SM-Higgs doublet. The covariant derivative is defined as
\begin{equation}
  \label{eq:cus-qdrpt:3}
  D_{\mu}\Theta =\partial_{\mu}\Theta + i g W_{\mu}\Theta - i g' B_{\mu} \Theta T_{3/2}^{3}.
\end{equation}
From Eq.~\eqref{eq:cus-qdrpt:1} we can derive the equations of motion for $\Theta$ and integrate it out at tree level. We find
\begin{align}
  \begin{aligned}
    \label{eq:cus-qdrpt:4}
    \Delta\mathcal{L}_{6}&=\frac{\lambda_{\Theta}^{2}}{\mu^{2}_{\Theta}}(\phi^{\dagger}\phi)^{3}\\
    \Delta\mathcal{L}_{8}&=\frac{\lambda_{\Theta}^{2}}{2\mu^{4}_{\Theta}}\left(5(\phi^{\dagger}\phi)^{2}(D_{\mu}\phi)^{\dagger}(D^{\mu}\phi)+(\phi^{\dagger}\phi)D_{\mu}(\phi^{\dagger}\phi)D^{\mu}(\phi^{\dagger}\phi)\right)-\frac{\lambda_{\Theta}^{2}}{\mu^{4}_{\Theta}}\left(\lambda'+\frac{15}{4}\widetilde{\lambda}\right)(\phi^{\dagger}\phi)^{4}.
  \end{aligned}
\end{align}
The contribution to $\Delta\mathcal{L}_{6}$ is consistent with \cite{deBlas:2014mba}. There, the contribution of $\Theta_{3}$ is $3 \lambda_{\Theta}^{2}/(2\mu^{2}_{\Theta}) $ and the one of $\Theta_{1}$ (with the relation $\lambda_{\Theta_{1}}=-\sqrt{3}\lambda_{\Theta_{3}}$ coming from Eq.~\eqref{eq:cus-qdrpt:1}) is also  $3\lambda_{\Theta}^{2}/(2\mu^{2}_{\Theta}) $. The resulting factor of $3$ is absorbed in the different definition of $\mathcal{O}_{6}$ compared to ours. 
We see that at dimension eight the model induces the desired contribution to the Higgs potential, as well as two more contributions with two derivatives. All of them conserve custodial symmetry. 

We have also checked, using \texttt{SARAH}~\cite{Staub:2013tta}, that loop corrections to the $\rho$ and $S$ parameters are well within the experimental bound.
The collider phenomenology of the custodial quadruplet can be understood in terms of the unbroken $SU(2)_V$. The Higgs bi-doublet decomposes as $(\mathbf{2}, \mathbf{2})= 1 + \mathbf{3}$, while the custodial bi-quadruplet decomposes as $(\mathbf{4},\mathbf{4}) = 1 + \mathbf{3} + \mathbf{5} + \mathbf{7}$. The latter singlet and triplet contain only electrically neutral and singly-charged scalars, which are difficult to produce and detect at colliders. Note that they only couple to the SM fermions via the mixing with the Higgs singlet and triplet. Moreover, this mixing is very small: After all, $\mathcal{O}_6$ is the only operator generated at tree level, which does not modify the Higgs couplings at low energy. This also suggests that measuring the Higgs couplings is not the most promising strategy to test this setup. 

Moreover, the septuplet contains large electric charges. However, these cannot directly decay into pairs of SM particles~\footnote{Note that there is no $SU(2)_V$ septuplet constructed out of two $1$ and/or two $\mathbf{3}$. The septuplet cannot even decay into three triplets: Although allowed by $SU(2)_V$, operators mediating this decay would contain at least three gauge bosons and one scalar, while Lorentz invariance forbids this kind of interaction at dimension four.}. They decay only via the emission of (soft) gauge bosons into lower-charged states in the custodial quadruplet, which are also difficult to test at colliders. The quintuplet, instead, can be both efficiently produced (in pairs via EW interactions) and decays mostly into pairs of gauge bosons 
(indeed $\mathbf{3}\times\mathbf{3} = 1 + \mathbf{3} + \mathbf{5}$). Decays into pairs of Higgs bosons are not allowed, because this is a complete singlet of $SU(2)_V$. In particular, the doubly-charged, singly-charged and neutral components of the quintuplet decay with branching ratios 
\begin{equation}
\text{Br}(\Theta^{\pm\pm}\to W^\pm W^\pm) = 1, \quad \text{Br}(\Theta^{\pm}\to W^\pm Z) = 1, \quad \text{Br}(\Theta^0\to W^+ W^- + ZZ) = 1~. 
\end{equation}

\begin{figure}[t]
\includegraphics[width=0.49\columnwidth]{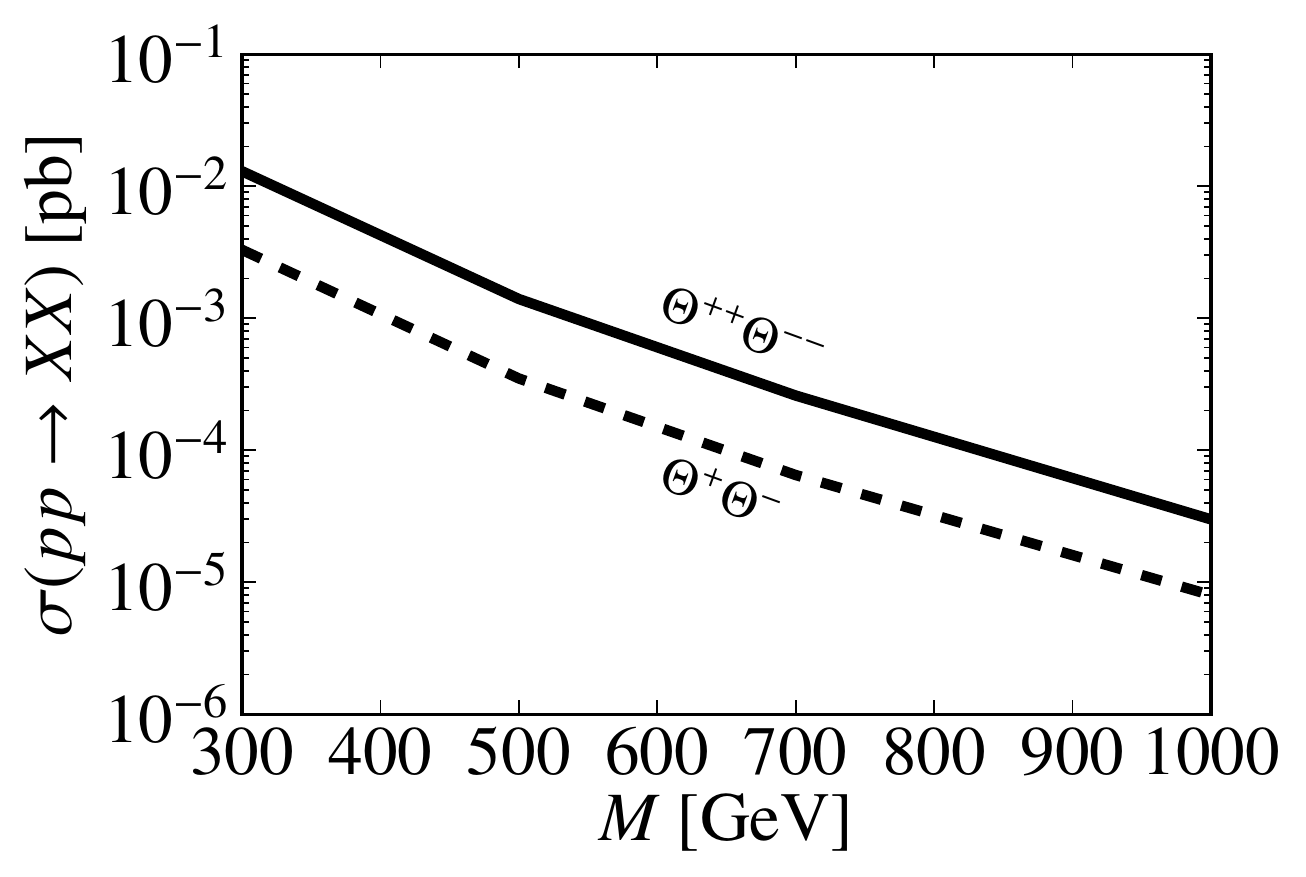}
\includegraphics[width=0.49\columnwidth]{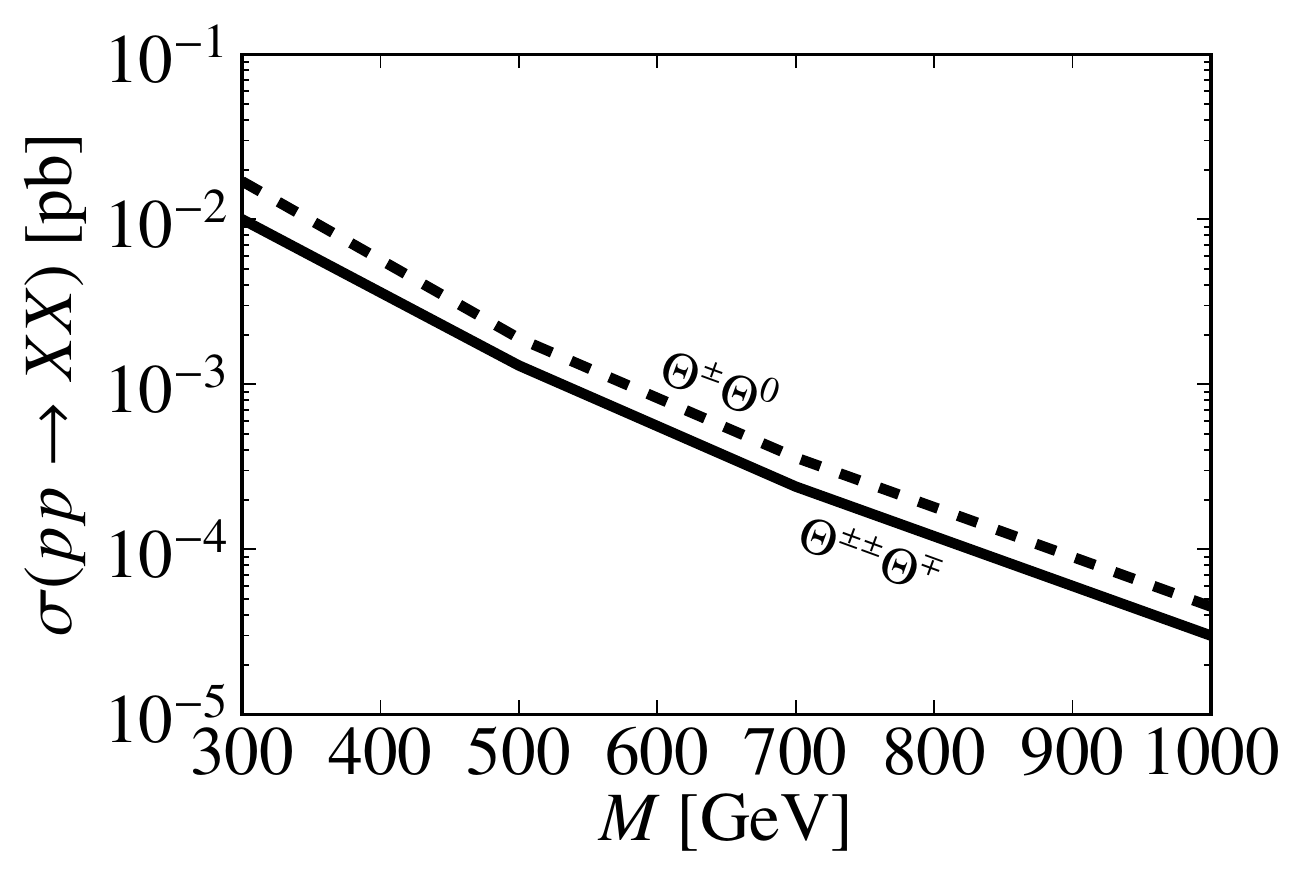}
\includegraphics[width=0.49\columnwidth]{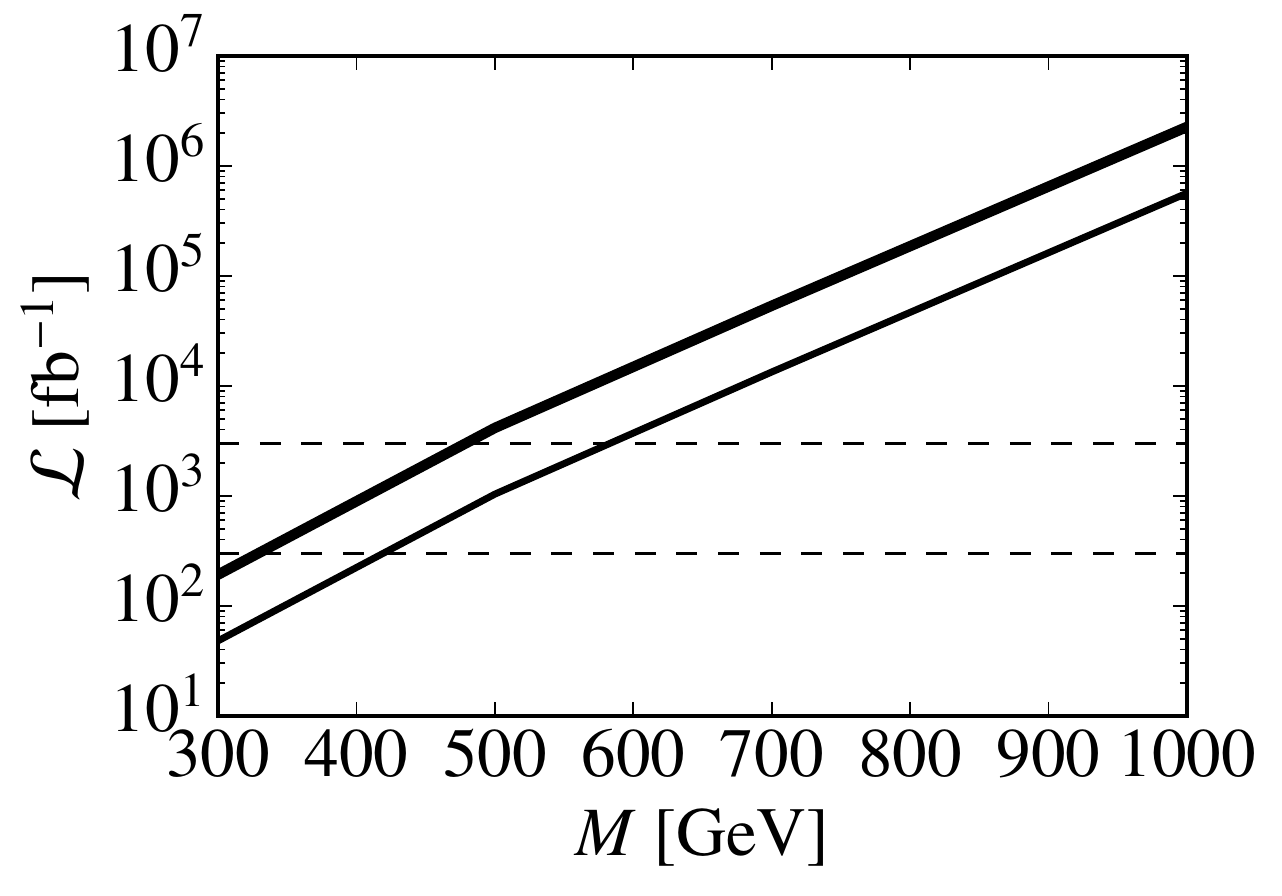}
\hspace{0.3cm}\includegraphics[width=0.49\columnwidth]{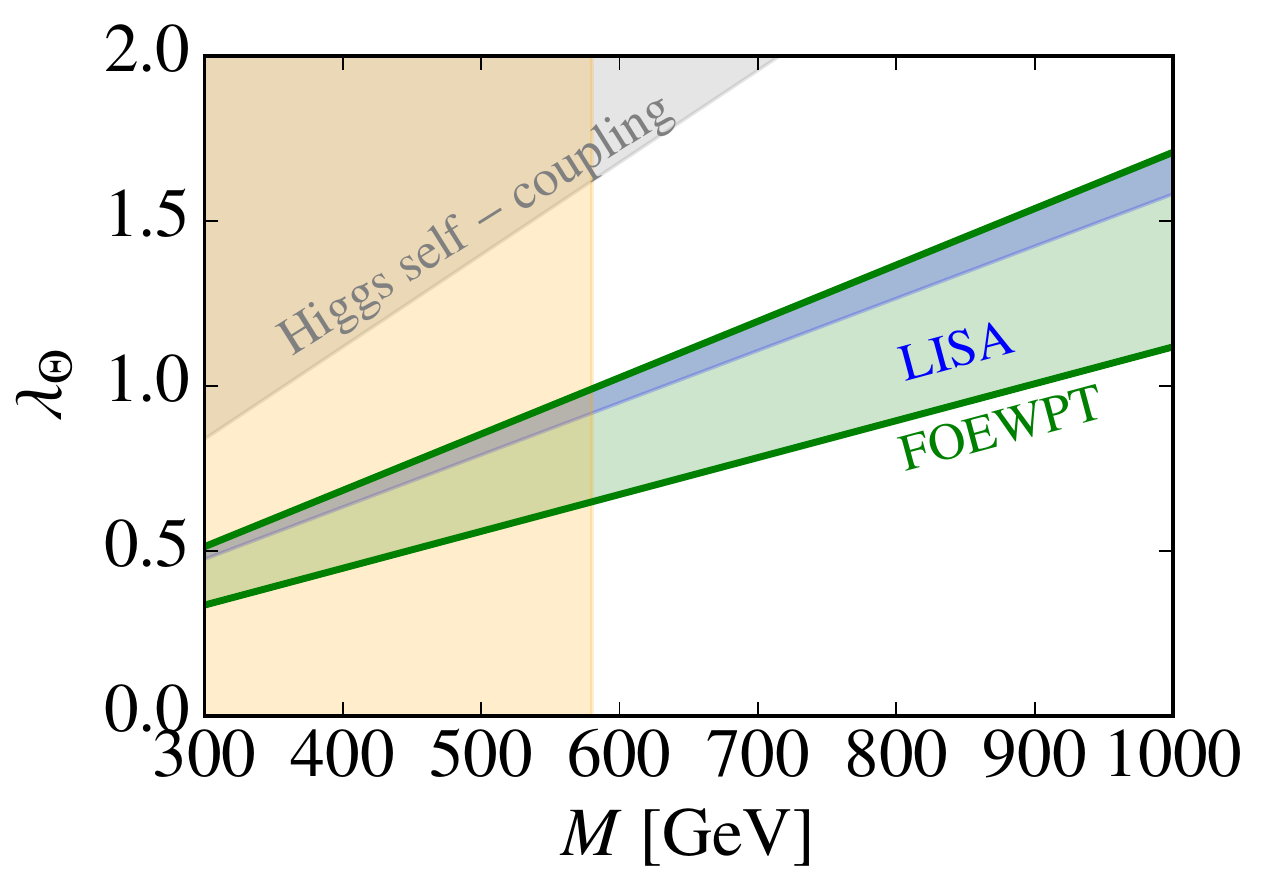}
\caption{\it Upper left) Neutral current cross sections for pair-production of scalars in the custodial quadruplet model. Upper right) Same as before but for the charged current. Bottom left) Integrated luminosity required to exclude the custodial quadruplet at the 95\% C.L. for different masses using two different analyses; see text for details. Two representative values of the collected luminosity, $\mathcal{L} = 300$ fb$^{-1}$ and $\mathcal{L} = 3$ ab$^{-1}$ are also shown with dashed lines. Bottom right) Parameter space region where the FOEWPT takes place for $\lambda^\prime =\tilde{\lambda}=1$ and the reach of different searches. The yellow region shows the HL-LHC reach taken from the bottom-left panel.}\label{fig:xsecs}
\end{figure}

We implement this model in \texttt{MadGraph v5}~\cite{Alwall:2014hca} by means of \texttt{Feynrules v2}~\cite{Alloul:2013bka}. We subsequently compute the pair-production cross sections mediated by neutral and charged currents for masses in between $300$ and $1000$ GeV. The results are shown in the upper left and upper right panels of Fig.~\ref{fig:xsecs}, respectively.

We have also estimated the current and the future LHC reach for this scenario. To this aim, we have generated Monte Carlo events, including radiation, fragmentation and hadronization effects with \texttt{Pythia v6}~\cite{Sjostrand:2006za}, and analysed them using \texttt{CheckMate v2}~\cite{Dercks:2016npn}. The latter implements several multi-lepton SUSY searches. Among them, the search that turns out to be the most sensitive to our scenario, is the ``$SR3\ell-H$" signal region of Ref.~\cite{ATLAS-CONF-2016-096}, which looks for three leptons, no $b$-jets and large missing energy. This analysis considers $13$ fb$^{-1}$ of LHC data at $\sqrt{s} = 13$ TeV. The integrated luminosity needed to exclude a particular value of the quadruplet mass at the  95\% C.L. can then be estimated as
\begin{equation}
 \mathcal{L} = 13~\text{fb}^{-1}\times \frac{1}{(s/s_{excl})^2}~,
\end{equation}
where $s/s_{excl}$ is the number of expected signal events over the number of excluded signal events as reported by \texttt{CheckMate}. The corresponding result is represented by the thick solid line in the left bottom panel of Fig.~\ref{fig:xsecs}. The thin solid line represents the luminosity required to test the different masses using the improved multi-lepton search described in Ref.~\cite{Alcaide:2017dcx}. As things stand, masses as large as $M\sim 600$ GeV can be tested in multi-lepton final states at the LHC. Getting ahead of the results discussed, we also show the reach of LHC Higgs-self couplings measurements as well as that of the gravitational wave observatory LISA; see right bottom  panel in Fig.~\ref{fig:xsecs}. Interestingly, the former cannot even test the parameter space region where the FOEWPT takes place. (As a matter of fact, in the present scenario the LHC Higgs-self couplings measurements are sensitive only to the region where the theory does not achieve EWSB.).
These results suggest that most weakly-coupled models (those containing $SU(2)_L$ charged states), even if tuned to avoid large corrections to operators other than $\mathcal{O}_6$, can be better tested at gravitational wave observatories or in direct LHC searches~\footnote{Note that most SM extensions avoiding large corrections to operators such as $\mathcal{O}_{\phi D}$ or $\mathcal{O}_{d6}$ involve different multiplets and therefore charged (often doubly-charged) scalars. One possible counter-example is a singlet scalar whose own parameters are tuned; see Ref.~\cite{DiVita:2017eyz}.}.
\subsection{Strongly-coupled models}
\label{sec:ewXL}
So far, we discussed the dynamics of the electroweak phase transition in presence of effective modifications of the scalar potential only, as well as potential UV-completions that lead to this particular pattern of low-energy effects. Working in a generic bottom-up EFT, we would in principle have many more effective operators, with coefficients of similar size to the coefficients that modify the potential. To 
overcome the strong experimental constraints 
on these operators, we require a hierarchy between the large effects in the scalar sector and the more constrained effects in the gauge-fermion sector. This can be achieved with a strongly-coupled UV-completion. While the complete description of such a UV-completion requires lattice simulations (and is therefore more model-dependent), we can describe the low-energy effects by assuming a mass gap between the (pseudo-) Nambu-Goldstone bosons and the higher resonances of the theory. 
The EW chiral Lagrangian (\ewXL) ~\cite{Feruglio:1992wf,Bagger:1993zf,Koulovassilopoulos:1993pw,Burgess:1999ha,Wang:2006im,Grinstein:2007iv,Contino:2010mh,Contino:2010rs,Azatov:2012bz,Alonso:2012px,Buchalla:2012qq,Buchalla:2013rka,Buchalla:2013eza} is the most general EFT that describes such low-energy effects of strongly-coupled new physics. Historically, it emerged from the Higgs-less chiral Lagrangian \cite{Appelquist:1980vg,Longhitano:1980tm,Dobado:1990zh,Herrero:1993nc}, which was then supplemented with a generic scalar singlet $h$. Since this does not assume any IR-doublet structure for the Higgs, it describes a very wide class of new-physics models that induce large deviations in the Higgs sector from the SM. The leading-order {\ewXL}  is %
\begin{align}
  \label{eq:s1}\nonumber
  L_{\text{LO}}^{ew\chi} =& -\frac{1}{2} \langle G_{\mu\nu}G^{\mu\nu}\rangle -\frac{1}{2}\langle W_{\mu\nu}W^{\mu\nu}\rangle -\frac{1}{4} B_{\mu\nu}B^{\mu\nu} \\\nonumber
  &+i\bar{q}_{L}\slashed{D}q_{L} +i\bar{\ell}_{L}\slashed{D}\ell_{L} +i\bar{u}_{R}\slashed{D}u_{R} +i\bar{d}_{R}\slashed{D}d_{R} +i\bar{e}_{R}\slashed{D}e_{R} \\\nonumber
  &+\frac{v^2}{4}\ \Tr{(D_\mu U^\dagger D^\mu U)} \left( 1+F(h)\right) +\frac{1}{2} \partial_\mu h \partial^\mu h - V(h)\\\nonumber
  &- \frac{v}{\sqrt{2}} \left[ \bar{q}_{L} \left( Y_u +\sum\limits^\infty_{n=1} Y^{(n)}_u \left(\frac{h}{v}\right)^n \right) U P_{+}q_{R} + \bar{q}_{L} \left( Y_d +  \sum\limits^\infty_{n=1} Y^{(n)}_d \left(\frac{h}{v}\right)^n \right) U P_{-}q_{R} \right.\\
  &\left. + \bar{\ell}_{L}  \left( Y_e +\sum\limits^\infty_{n=1} Y^{(n)}_e \left(\frac{h}{v}\right)^n \right) U P_{-}\ell_{R} + \text{ h.c.}\right] ~,
\end{align}
where $U$ stands for the exponential of the Goldstone matrix,
$G, W$ and $B$ are the SM gauge fields, $u_R$, $d_R$, $e_R$, $q_L$ and $\ell_L$ are the fermions of the SM, and $Y$ are generalised Yukawa couplings. The scalar $h$ couples through general polynomials to the other fields, which reflects its strongly-coupled origin. 

These polynomials ($V(h), F(h)$, and $Y_i(h)=Y_i+\sum_{n=1}^\infty Y_i^{(n)} (h/v)^n)$ are not 
truncated at canonical dimension four, but go to arbitrary order.
(An additional operator of the structure $(\partial_{\mu}h)(\partial^{\mu})f(h)$ is also allowed by symmetry, but can be removed via field redefinitions, without loss of generality~\cite{Buchalla:2013rka}.) The coefficients of these polynomials depend on $v/f$.

As the Lagrangian in Eq.~\eqref{eq:s1} contains terms with arbitrarily high canonical dimension, the EFT can clearly not be organized in terms of canonical dimensions. Instead, it is organised by a generalisation of the momentum expansion of chiral perturbation theory~\cite{Weinberg:1978kz}, the chiral dimensions~\cite{Buchalla:2013rka,Buchalla:2013eza}. They reflect an expansion in terms of loops, which guarantees the renormalizability of the EFT at a fixed order in the expansion.
The cutoff of the EFT is at $\Lambda=4\pi f$, yielding the expansion parameter $f^{2}/\Lambda^{2} = 1/ 16\pi^{2}$. For $v < f$, the parameter $\xi = v^{2}/f^{2}$ is smaller than the unity and Eq.~\eqref{eq:s1} can be further expanded in $\xi$.
In this scenario, a double expansion in $\xi$ and $1/16\pi^{2}$ organises the EFT~\cite{Buchalla:2014eca}, in the spirit of the strongly-interacting light Higgs Lagrangian \cite{Giudice:2007fh}. \\
In this double expansion, we still see some of the decoupling effects, but also a pattern of Wilson coefficients that is coming from the strong sector. Depending on the structure of the operators, they will be suppressed by ratios
of scales ($\xi$, based on their canonical dimension) and loop factors ($1/16\pi^{2}$, based on their chiral dimension). This creates an additional hierarchy among the operators of a given canonical dimension, compared to the weakly-coupled case of section \ref{sec:weakly}. Some of the dimension six operators, corresponding to $L_{\text{LO}}^{ew\chi}$, will only be suppressed by $\xi$, while other operators, corresponding to $L_{\text{NLO}}^{ew\chi}$, will be suppressed by an additional loop factor, resulting in $\xi/16\pi^{2}$. The former affects the Higgs sector with deviations of $\mathcal{O}(10\%)$, dominating over effects in the gauge-fermion sector of the latter group, with deviations of $\mathcal{O}(1\%)$ or below. This hierarchy also reflects the current experimental constraints: The gauge-fermion sector is rather strongly constrained, while large effects in the Higgs couplings are still possible.
The \ewXL\ of Eq.~\eqref{eq:s1} is now expanded in both chiral and canonical dimensions.

Since $\xi=\mathcal{O}(0.1 - 0.2)$~\cite{Khachatryan:2016vau,Buchalla:2015qju,Sanz:2017tco,deBlas:2018tjm}, effects of $\mathcal{O}(\xi^{2})$ could in principle be larger than the $\mathcal{O}(1/16\pi^{2})$ effects.
The leading effects in the double expansion are then given by expanding $L_{\text{LO}}$ up to $\mathcal{O}(\xi^{2})$. \textit{A priori}, the Higgs potential, which at this order contains both $\mathcal{O}_6$ and $\mathcal{O}_8$, is of chiral dimension $0$ and the dominating effect. However, the Higgs mass is then expected to be of order $\mathcal{O}(\Lambda)$, which would break the EFT approach. In order for this to make sense, the Higgs mass must be parametrically suppressed to appear at chiral dimension of 2~\footnote{This occurs naturally in composite Higgs models (CHMs), where the Higgs potential is generated radiatively and then comes with two powers of weak couplings ($g^2, y^2$) and a corresponding loop suppression of the scale $\Lambda^2$ to the scale $f^{2}$.}. An additional fine tuning of $\mathcal{O}(\xi)$ is needed for $m_{h}\sim v$. This, however, might only affect the mass term of the potential and the Higgs self-couplings could have large deviations from the SM, induced by $c_6$ and $c_8$.

We can understand the enhancement on the operators in the potential by just dimensional analysis if we assume that the strongly-coupled theory is described by only one relevant coupling $g_\ast$. To this aim, we need to abandon the convention $\hbar = c = 1$ recovering the physical dimensions of these two constants. It turns out that the coefficient of any operator involving $r$ Higgs insertions and $q$ derivatives scales as $g_\ast^2 f^4 [h/f]^r [\partial/(g_\ast f)]^q$, up to $\mathcal{O}(1)$ coefficients~\cite{Giudice:2007fh,Panico:2015jxa,Buchalla:2016sop,Chala:2017sjk}. Hence, scalar operators not carrying derivatives are enhanced with respect to the derivative ones by several powers of $g_\ast$ ($ \gg 1$ in a strongly couple theory); \textit{e.g.} $c_6\sim g_\ast^2$ versus $c_{d6}\sim 1$. We refer to Ref.~\cite{Azatov:2015oxa} for a discussion on which scenarios show this enhancement while still having $m_h \sim v$. This justifies why we studied the effects of $\mathcal{O}_{6}$ and $\mathcal{O}_{8}$, neglecting other effects, as first approximation. 

To account for all leading effects consistently, we have to consider the full set of dimension-six and dimension-eight operators that contribute at chiral dimension $2$ for the expansion in $\xi$. The operators are
\begin{align}
  \label{eq:d-e:1}\nonumber
  \left(\phi^\dagger\phi\right)^{3}, \quad \partial_{\mu}\left(\phi^\dagger\phi\right)\partial^{\mu}\left(\phi^\dagger\phi\right), \quad \bar\Psi Y \phi\Psi\left(\phi^\dagger\phi\right), \\
  \left(\phi^\dagger\phi\right)^{4}, \quad \partial_{\mu}\left(\phi^\dagger\phi\right)\partial^{\mu}\left(\phi^\dagger\phi\right)\left(\phi^\dagger\phi\right), \quad \bar\Psi Y \phi\Psi \left(\phi^\dagger\phi\right)^{2}.
\end{align}
With the identification $\phi = \tfrac{(v+h)}{\sqrt{2}}\, U \binom{0}{1}$, we find at the different orders of $\xi$:
  \begin{align}
      \label{eq:d-e:2}\nonumber
      L_{\xi^{0}} &= \frac{1}{2}\partial_{\mu}h\partial^{\mu}h + \frac{\mu^{2}}{2}(v+h)^{2}-\frac{\lambda}{4}(v+h)^{4}-\tfrac{1}{\sqrt{2}}\bar\Psi\hat{Y}_{\Psi}U P_{\pm}\Psi (v+h)+\frac{v^2}{4}\ \Tr{(D_\mu U^\dagger D^\mu U)} \left( 1+\tfrac{h}{v}\right)^{2},\\\nonumber
      L_{\xi^{1}} &= \frac{c_{d6}}{2f^{2}}\partial_{\mu}h\partial^{\mu}h(v+h)^{2} -\frac{c_{6}}{8f^{2}}(v+h)^{6}-\tfrac{1}{\sqrt{2}f^{2}}\bar\Psi\hat{Y}_{\Psi}^{(6)}U P_{\pm}\Psi (v+h)^{3},\\
       L_{\xi^{2}} &= \frac{c_{d8}}{2f^{4}}\partial_{\mu}h\partial^{\mu}h(v+h)^{4} -\frac{c_{8}}{16f^{4}}(v+h)^{8}-\tfrac{1}{\sqrt{2}f^{4}}\bar\Psi\hat{Y}_{\Psi}^{(8)}U P_{\pm}\Psi (v+h)^{5}.
  \end{align}
To bring the Lagrangian to the form of $L_{\text{LO}}^{ew\chi}$ in Eq.~\eqref{eq:s1}, we have to canonically normalise the field $h$ using the field redefinition discussed in Ref.~\cite{Buchalla:2013rka}. We find \cite{Buchalla:2016bse}
\begin{align}
  \label{eq:d-e:3}\nonumber
  h\rightarrow h \Big\lbrace 1-\tfrac{\xi}{2} c_{d6} &\left(1+\tfrac{h}{v}+\tfrac{h^2}{3 v^2}\right)+\xi ^2 c_{d6}^2 \left(\tfrac{3}{8}+\tfrac{h}{v}+\tfrac{13}{12} \left(\tfrac{h}{v}\right)^2+\tfrac{13}{24} \left(\tfrac{h}{v}\right)^3+\tfrac{13}{120} \left(\tfrac{h}{v}\right)^4\right) \\ 
&  -\xi ^2 c_{d8} \left(\tfrac{1}{2}+\tfrac{h}{v}+\left(\tfrac{h}{v}\right)^2 +\tfrac{1}{2} \left(\tfrac{h}{v}\right)^3+\tfrac{1}{10} \left(\tfrac{h}{v}\right)^4\right)\Big\rbrace .
\end{align}
To obtain the right Higgs VEV and mass, the parameters $\mu^{2}$ and $\lambda$ have to fulfil
\begin{align}
\begin{aligned}
  \label{eq:d-e:4}
  \mu^{2} &= \frac{m_h^2}{2} +\frac{v^{2}}{f^{2}}\left(\frac{1}{2}c_{d6} m_h^2-\frac{3}{4} c_6v^{2}\right)+\frac{v^{4}}{f^{4}}\left(\frac{1}{2}c_{d8} m_{h}^{2}-c_8 v^{2}\right),\\
  \lambda&= \frac{m_{h}^{2}}{2 v^{2}}+\frac{v^{2}}{f^{2}}\left(\frac{c_{d6}}{2}\frac{m_{h}^{2}}{v^{2}}-\frac{3c_{6}}{2}\right)+\frac{v^{4}}{f^{4}}\left(\frac{c_{d8}}{2}\frac{m_{h}^{2}}{v^{2}}-\frac{3c_{8}}{2}\right).
\end{aligned}
\end{align}
Applying Eq.~\eqref{eq:d-e:3} everywhere in Eq. \eqref{eq:d-e:2}, we find the expansion of $V(h), F(h)$, and $Y(h)$ in $\xi$. 
Writing
\begin{align}
  \begin{aligned}
    \label{eq:d-e:5}
    V(h) &= \frac{1}{2}m_{h}^{2}v^{2}\left[\frac{h^{2}}{v^{2}}+\sum_{i=3}^{8}\lambda_{i}\left(\frac{h}{v}\right)^{i}\right],\\
    F(h) &= \sum_{i=1}^{6}f_{i}\left(\frac{h}{v}\right)^{i},
  \end{aligned}
\end{align}
we finally have
\begin{align}
  \begin{aligned}
  \label{eq:d-e:6}
  \lambda_{3}&=1+ \frac{v^{2}}{f^{2}} \left(2 c_{6}\frac{v^{2}}{m_{h}^{2}}-\frac{3}{2}c_{d6}\right)+\frac{v^{4}}{f^{4}}\left(\frac{15}{8}c_{d6}^{2}-3\frac{v^{2}}{m_{h}^{2}}c_{6}c_{d6}-\frac{5}{2}c_{d8}+4\frac{v^{2}}{m_{h}^{2}}c_{8}\right) ,\\
  \lambda_{4}&=\frac{1}{4} + \frac{v^{2}}{f^{2}}\left(3 c_{6}\frac{v^{2}}{m_{h}^{2}}-\frac{25}{12}c_{d6} \right)+\frac{v^{4}}{f^{4}}\left(\frac{11}{2}c_{d6}^{2}-9\frac{v^{2}}{m_{h}^{2}}c_{6}c_{d6}-\frac{21}{4}c_{d8}+8\frac{v^{2}}{m_{h}^{2}}c_{8}\right) ,
\end{aligned}
\end{align}
\begin{align}
  \begin{aligned}
    \label{eq:d-e:13}
    f_{1}&=2 - \frac{v^{2}}{f^{2}}c_{d6}+\frac{v^{4}}{f^{4}}\left(\frac{3}{4}c_{d6}^{2}-c_{d8}\right) ,\\
    f_{2}&=1 - 2\frac{v^{2}}{f^{2}}c_{d6}+\frac{v^{4}}{f^{4}}\left(3c_{d6}^{2}-3c_{d8}\right)
  \end{aligned}
\end{align}
and
\begin{align}
\begin{aligned}
\label{eq:d-e:7}
  Y_{\Psi}^{(1)} & =Y_{\Psi}+\frac{v^{2}}{f^{2}}\left(2\hat{Y}_{\Psi}^{(6)}-\frac{c_{d6}}{2}Y_{\Psi}\right)+\frac{v^{4}}{f^{4}}\left(4\hat{Y}_{\Psi}^{(8)}-\frac{c_{d8}}{2}Y_{\Psi}-c_{d6}\hat{Y}_{\Psi}^{(6)}+\frac{3}{8}c_{d6}^{2}Y_{\Psi}\right) ,\\
  Y_{\Psi}^{(2)} & =\frac{v^{2}}{f^{2}}\left(3\hat{Y}_{\Psi}^{(6)}-\frac{c_{d6}}{2}Y_{\Psi}\right)+ \frac{v^{4}}{f^{4}}\left(10\hat{Y}_{\Psi}^{(8)}-c_{d8}Y_{\Psi}-4c_{d6}\hat{Y}_{\Psi}^{(6)}+c_{d6}^{2}Y_{\Psi}\right),
  \end{aligned}
\end{align}
where we only list the couplings relevant for the subsequent discussion. The matrices $Y_{\Psi}$ and $ Y_{\Psi}^{(n)}$ are the fermion mass and Yukawa matrices defined in Eq. \eqref{eq:s1}. Note that the functional dependence of Eqs. \eqref{eq:d-e:6} and \eqref{eq:d-e:7} on $c_{i}$ differ from the result of Refs.~\cite{Buchalla:2014eca,Buchalla:2016bse}, as we do not include explicit factors of $\lambda$ in the definition of the Wilson coefficients. Already now we see two of the implications of adding these effective operators. The triple- and quartic-Higgs couplings are further modified with respect to the SM. 
Moreover, new vertices, such as $\bar\Psi \Psi hh$, also relevant for the study of double-Higgs production, arise.

Additionally, for current Higgs observables, also the local $GGh$ and $\gamma\gamma h$ operators are important, even though they are formally of next-to-leading order. This is because these amplitudes arise at the one-loop level of the leading-order Lagrangian; see Ref.~\cite{Buchalla:2015wfa}. Such a Lagrangian is 
\begin{equation}
  \label{eq:d-e:20}
  L_{GGh}=L_{kin}+G_{\mu\nu}G^{\mu\nu}\left[\frac{c_{g6}}{16\pi^{2}f^{2}}\phi^{\dagger}\phi+\frac{c_{g8}}{16\pi^{2}f^{4}}(\phi^{\dagger}\phi)^{2}\right]. 
\end{equation}
After symmetry breaking and the field redefinition of Eq. \eqref{eq:d-e:3}, this creates a contribution that renormalizes the gluon kinetic term and therefore $G_{\mu\nu}$. After this renormalization, we find
\begin{equation}
  \label{eq:d-e:21}
  L_{GGh}=G_{\mu\nu}G^{\mu\nu} \left[1+f_{G1}\frac{h}{v}+f_{G2}\frac{h^{2}}{v^{2}} +\mathcal{O}(h^{3})\right],
\end{equation}
with
\begin{align}
  \begin{aligned}
    \label{eq:d-e:22}
    16\pi^{2} f_{G1} &= \xi c_{g6} + \xi^{2}\left(c_{g8}-\frac{1}{2}c_{d6}c_{g6}-\frac{c_{g6}^{2}}{32\pi^{2}}\right),\\
    32\pi^{2} f_{G2} &= \xi c_{g6} +\xi^{2}\left(3 c_{g8}-\frac{1}{2}c_{d6}c_{g6}-\frac{c_{g6}^{2}}{32\pi^{2}}\right).
  \end{aligned}
\end{align}
The last term in each of the $f_{Gi}$ comes from the renormalization and is sub-leading. Finally, it is also worth noting that
all these operators would contribute to the EWPT, as they alter the $h_c$-dependent squared masses $m_i^2$ in Eqs.~\eqref{eq:masses}-\eqref{eq:masses2}. In addition, the derivative operator $\mathcal{O}_{d6}$ requires a reevaluation of the Coleman-Weinberg effective potential at finite temperature, as the field redefinition of Eq.~\eqref{eq:d-e:3} cannot be done in the unbroken phase~\cite{Burgess:2010zq,Passarino:2016pzb}. All these effects would be suppressed by $v^{2}/f^{2}$ in $a_{0}$, but would nevertheless have an impact on the computation of the quantities of the EWPT. 

Current experimental results only constrain effective couplings with a single Higgs field~\cite{Buchalla:2015qju,deBlas:2018tjm}, namely $Y_{t,b,\tau}^{(1)}, f_{1},$ and $f_{G1}$ from the list above. From these, $f_{1}$ is the most constrained, but still allows for deviations of $\mathcal{O}(5\%)$. The others are not constrained beyond $\mathcal{O}(10\%)$. While from a bottom-up point of view a deviation in one of these couplings might hint to a deviation in $\lambda_{3}$ of comparable size, such conclusions are strongly model dependent.\\
Double Higgs production, which would shed light on the $\lambda_{3}$ coupling of the Higgs potential in the SM, depends on five of the effective parameters from above~\cite{Grober:2015cwa,Kim:2018uty} if we restrict ourselves to the top loops only. These are $Y_{t}^{(1)}, Y_{t}^{(2)}, f_{G1}, f_{G2},$ and $\lambda_{3}$. A large deviation in $\lambda_{3}$ from its SM value could then be not seen in the experiment because of the interplay with the otherwise unconstrained other parameters. \\

\section{Conclusions}
\label{sec:conclusions}

It is well known that the presence of higher-dimensional operators in the Standard Model Higgs potential can drastically influence the dynamics of the ElectroWeak (EW) symmetry breaking. Among the possible operators, the interactions $\mathcal O_n=(\phi^\dagger\phi)^{\frac{n}{2}}$, with $\phi$ being the Higgs doublet, have attracted a lot of attention to make the EW Phase Transition (EWPT) strongly first order while evading any scheduled LHC search.
Achieving a strongly first order 
EWPT requires $c_6/f^2\gtrsim 1$ TeV$^{-2}$, with $f$ the cutoff of the theory and $c_6$ the coefficient of $\mathcal O_6$. This implies that $f$ is likely too close to the EW scale for the dimension-six EFT to be accurate, at least in weakly-coupled theories. Dimension-eight operators have then to be considered as well, which is also the case when strongly-coupled sectors are present. Such sectors
can also lead to naturally large corrections to the Higgs potential (in comparison with other operators). In view of this possibility, we have also examined the EFT where (only) both $\mathcal O_6$ and $\mathcal O_8$ are unsuppressed.

In the aforementioned dimension-eight setup, we have computed the parameters relevant for the EWPT, including the critical and nucleation temperatures and the %
VEVs of the Higgs at these temperatures. We have also estimated the latent heat and the inverse duration time of the phase transition, characterising the gravitational waves produced in the collisions of nucleated bubbles. Regarding the coefficients of $\mathcal{O}_6$ and $\mathcal{O}_8$, $c_6$ and $c_8$ respectively, we have obtained that the parameter region $3 \lesssim c_6/f^2 + 3 v^2 c_8/(2 f^4) \lesssim 3.5$ is in the reach of the future LISA experiment. Remarkably, due to the low LHC sensitivity to $\mathcal{O}_6$ and $\mathcal{O}_8$, LISA will be the first experiment able to significantly constrain these operators. Concerning the reach of future colliders, we have shown that almost all values of interest will be probed by a future FCC-ee in double-Higgs production, while the whole parameter space will be testable combining double- and triple-Higgs production in hadronic colliders.

Given that the new physics matching the previous EFT must be 
quite low, we have also explored the possibility of producing the supposely heavy new fields
at the LHC. Among the ultraviolet completions exhibiting only the operators $\mathcal O_n$, we have proven that in weakly-coupled setups consisting of new scalar singlets or triplets, the presence at low energies of other effective operators already quite constrained by LHC and EW precision data is unavoidable. (Of course, in scenarios with several scalars, a tuning in the fundamental parameters can still yield to an EFT where the coefficients 
$c_6$ and $c_8$ are substantially larger than those of the other effective operators.) On the contrary, in models involving only doublets or quadruplets (higher representations do only lead to $\mathcal{O}_6$ at the loop level, being $c_6$ therefore very small to modify the EWPT), new symmetries can make all operators other than those modifing the scalar potential vanish. Such models still contain charged particles
that can be produced in pairs via Drell-Yan and then decay into  longitudinal polarizations of the gauge bosons. 
We have shown that even in the particular case of a custodial quadruplet, the LHC reach is far smaller than that of LISA.
%

\acknowledgments
We are grateful to Florian Staub for useful support on \texttt{SARAH}. MC acknowledges AEC for hospitality during the first state of this project. CK thanks Andrew J.~Long for discussions about the effective potential, and the Enrico Fermi Institute at the University of Chicago and Fermi National Laboratory for hospitality, where parts of this research was carried out. CK acknowledges also fruitful discussions at the LHCPHENO 2017 workshop at IFIC Valencia. MC is supported by the Royal Society under the Newton International Fellowships programm. The work of CK is supported in part by the Spanish Government and ERDF funds from the EU Commission [Grants No.~FPA2014-53631-C2-1-P and SEV-2014-0398], by the Alexander von Humboldt-Foundation, and by the Fermi Research Alliance, LLC under Contract No.~DE-AC02-07CH11359 with the U.S.~Department of Energy, Office of Science, Office of High Energy Physics.
 GN is financed by the Swiss National Science Foundation (SNF) under grant 200020-168988. 
%

\bibliographystyle{JHEP}
\bibliography{references}{}

\end{document}